\documentclass[letterpaper,fleqn,11pt]{article}

\usepackage{graphicx,amsmath}
\usepackage[text={6.5in,9in},centering]{geometry}

\newcommand{\ms}{\scriptscriptstyle\sf}
\DeclareMathOperator{\sech}{sech}
\newtheorem{theorem}{Theorem}[section]

\begin{document}
\noindent
\begin{center}
{\Large\bf Mass Transport Phenomena in a MCFC Cathode}
\end{center}

\vspace{0.5cm}

\begin{center}
Peter {\sc Berg}$^1$ and Justin {\sc Findlay}

\vspace{0.5cm}

Faculty of Science, University of Ontario
Institute of Technology, \\
2000 Simcoe Street N., Oshawa, ON, L1H 7K4, Canada, \\
$^1$Email: peter.berg@uoit.ca, Phone: +1 905 721 8668,\\ 
FAX: +1 905 721 3304, web: www.peterberg.net 

\vspace{0.5cm}

05 August 2009

\vspace{1.5cm}

{\bf Abstract}
\end{center}
A molten carbonate fuel cell (MCFC) is an electro-chemical energy
conversion technology that runs on natural gas and employs a molten
salt electrolyte.  In order to keep the electrolyte in this state, the
cell must be kept at a temperature above $500^{\circ}\rm{C}$,
eliminating the need for precious metals as the catalyst.  There has
been only a limited amount of research on modelling the transport
processes inside this device, mainly due to its restricted
applicability for mobile applications. 

In this work, three one-dimensional models of a MCFC cathode are
presented based on different types of diffusion and
convection. Comparisons between models are performed so as to assess
their validity. Regarding ion transport, it is shown that there exists
a limiting case for ion migration across the cathode that depends on
the conductivity for the liquid potential.  Finally, an optimization
of the diffusivity across the cathode is carried out in an attempt to
increase the cell performance and its longevity.  

\vspace{0.5cm}

\cleardoublepage

{\bf Nomenclature}
\begin{table}[h]
  \begin{center}
    \begin{tabular}{lllc}
      $b$ & Bruggeman correction, & $1.5$ & \cite{bruggeman:35a,sole,white:2003b} \\
      $c$ & Concentration, $\rm{mol/m^3}$ & & \\
      $c_{\ms T}$ & Total concentration, $\rm{mol/m^3}$ & & \\
      $D$ & Diffusivity ($\sf{O_2}$ in air), $\rm{m^2/s}$ & $2.5\times10^{-5}$ & \\
      %$E$ & Actual cell potential, $\rm{V}$ & & \\
      %$E^{ohmic}$ & Ohmic polarization, $\rm{V}$ & & \\
      %$E_r$ & Reversible cell potential, $\rm{V}$ & & \\
      $F$ & Faraday's constant, $\rm{C/mol}$ & $96487$ & \\
      %$\rm{I}$ & Current, $\rm{A}$ & & \\
      $i_0$ & Exchange current density, $\rm{A/m^2}$ & $1\times10^{-3}$ & \\      
      %J$ & Flux, & & \\
      %J_{\rm I}$ & Current density, $\rm{A/m^2}$ & & \\
      $L$ & Thickness of cathode, $\rm{m}$ & $8\times10^{-4}$ &
      \cite{white:2003b} \\
      $n$ & Number of electrons in cathode reaction, & $4$ & \\
      $P$& Pressure & & \\
      %$\rm{R}$ & Resistance, $\rm{\Omega}$ & & \\
      $R$ & Ideal gas constant, $\rm{J/K\,mol}$  & $8.314$ & \\
      $T$ & Temperature, $\rm{K}$ & $900$ & \cite{li:06} \\
      $u$ & Fluid velocity, $\rm{m/s^2}$ & & \\
      %$\rm{V}$ & Voltage, $\rm{V}$ & & \\
       & & & \\
      $\alpha$ & Transfer coefficient, & $0.5$ & \cite{white:2003b} \\
      $\epsilon_g$ & Gas porosity, & $0.4$ & \cite{white:2003b} \\
      $\epsilon_l$ & Liquid porosity, & $0.3$ & \cite{white:2003b} \\
      $\epsilon_s$ & Solid porosity, & $0.3$ & \cite{white:2003b} \\
      %$\eta_a$ & Activation polarization, $\rm{V}$ & & \\
      $\eta$ & Polarization coefficient, $\rm{V}$ & & \\      
      $\kappa$ & Permeability, $\rm{m^2}$ & $1.9\times10^{-12}$ &
      \cite{promislow} \\
      $\mu$ & Viscosity, $\rm{kg/m\,s}$ & $2.25\times10^{-5}$ &
      \cite{promislow} \\      
      $\nu$ & Stoichiometric coefficient & & \\
      $\sigma_l$ & Liquid conductivity, $\rm{S/m}$ & $140$ &
      \cite{fehribach:96} \\
      $\sigma_s$ &   Solid conductivity, $\rm{S/m}$ & $1300$ &
      \cite{white:2003b}\\
      $\phi$ & Potential, $\rm{V}$ & & \\
    \end{tabular}
  \end{center}
%  \caption{List of standard parameters used in model simulations.}
  \label{standards}
\end{table}
\cleardoublepage

\section{Introduction}
Molten Carbonate Fuel Cells (MCFCs) are widely being considered for
stationary power generation and a better understanding of the
transport processes in the electrodes and cells are needed to improve
viability.  Recently, a MCFC system was built at Enbridge headquarters
in Toronto, Ontario.  Their system features four separate MCFC stacks
where the primary fuel, natural gas, is pumped into the system from a
pressure let-down station.  The excess heat released by the fuel cell
system is used to heat the adjacent building and preheat the gas
before expansion.

Molten carbonate fuel cells were initially developed with the
intention of operating directly on coal.  The primary fuel currently
in use is either coal-derived gases or more commonly natural gas
\cite{li:06}.  MCFCs are still under development and have not reached
market acceptance as a possible primary or secondary source of
energy.

The concept of the MCFC is almost a century old with the first patent
awarded in 1916 to W.D. Treadwell.  It was first conceived in Europe
in the 1940s as an attempt to convert coal to electricity in carbonate
media.  An initial demonstration was successfully completed by Broers
and Ketelaar in the 1950s, with the first pressurized stack appearing
in the 1980s.  Most of our current knowledge stems from work done in
the 1970s and 1980s~\cite{li:06}.

Current development concentrates on base-load utility applications as
well as dispersed or distributed electric-power generation with heat
co-generation.  Due to low power densities and long start-up times,
there is limited potential for mobile applications, although MCFCs might
be suitable for power-trains for large surface ships and trains.  

This work presents three mass transport models of a MCFC cathode
electrode to be compared for the same parameters.  The second section
of this work provides a brief introduction to the physical and
chemical processes in the cathode electrode.  The third section
focuses on the three models being used as well as numerical results.
Section 4 presents an analytical resolution to the non-existence of
numerical solutions found by altering the liquid conductivity
parameter based on values in White {\em et al}.~\cite{white:2003b}.
Section 5 presents an optimization model for the mass transport of a
single species in the MCFC cathode electrode. Conclusions will be
drawn based on numerical and analytical results in the fifth chapter.
Before we begin our analysis, a brief overview of a MCFC cathode is
provided and models based on mass transport phenomena shall be described.

\section{Cathode}
Inside the cathode, oxygen and carbon dioxide flow in the same
direction across the electrode at different rates.  The three-phase
boundary allows for reactions to occur along the length of the domain
of the electrode.  Across the channel interface, only oxygen and
carbon dioxide are able to flow while the liquid electrolyte is kept from
flowing into the channel causing corrosion.  In fact, the liquid
electrolyte distribution is only controlled by capillary pressure.
Therefore, the pore size must be maintained very carefully during
manufacturing.  At the electrode/electrolyte boundary, the electrolyte
concentration is much larger and fills the pores, which helps to avoid
any gas leakage into the electrolyte assembly.  

The inlet gas at the cathode is mainly composed of $\sf{N_2}$,
$\sf{O_2}$ and $\sf{CO_2}$ but contains trace amounts of other
molecules since the main source of oxygen comes from air.  The effects
of nitrogen are neglected in this thesis since it does not flow within
the cathode.  Only a few metals are stable as a cathode material due
to the extremely corrosive nature of the molten carbonate electrolyte
and currently nickel oxide is in use.  Only semiconducting oxides are
feasible from a cost point of view~\cite{li:06}.  The mean pore size
of $\sf{NiO}$ electrodes is about $10\,\rm{\mu}\rm{m}$.  The smaller
pores are filled with the electrolyte to form the three-phase boundary
needed for the reaction, while the larger pores remain open for gas
flow. Nickel oxide is also slightly soluble in the electrolyte which
limits the lifetime of the cell, according to
\begin{equation}
\sf{NiO+CO_2\to Ni^{2+}+CO_3^=}.
\end{equation}
The optimal cathode performance depends upon the gas composition where
there exists a 2:1 ratio of $\sf{CO_2}$ to $\sf{O_2}$ consumed in the overall
electro-chemical reaction.  In order to reduce the $\sf{NiO}$ solubility and
increase lifetime, the carbon dioxide concentration should be reduced,
although if it is too low, the dissociation of carbonate ions becomes
significant
\begin{equation}
\sf{CO_3^=\to CO_2+O^=},
\end{equation}
thereby limiting the cell lifetime due to electrolyte losses.  The balance
between $\sf{NiO}$ solubility and dissociation of $\sf{CO_3^=}$ can become very
difficult to control, and to predict cell lifetime is generally challenging.

The optimal thickness of the electrode, which depends upon the gas
composition and current density as well as other operating conditions,
ranges from $0.4-0.8\,\rm{mm}$~\cite{li:06}.

At the three-phase boundary in the cathode, oxygen and carbon dioxide
diffuse towards the electrolyte which has penetrated the $\sf{NiO}$
pore. Where the gas flow meets the electrolyte, the gas molecules are
absorbed into the electrolyte, react with the electrons at the surface
of the electrode, and produce carbonate ions.  The electro-chemical
reaction is given by
\begin{equation}
\sf{\frac{1}{2}O_2+CO_2+2e^-\to CO_3^=}.
\label{cathode_react}
\end{equation}

\section{Mathematical Models of Diffusion}
\label{models}
\begin{figure}[t]
  \begin{center}
  \includegraphics[width=65mm]{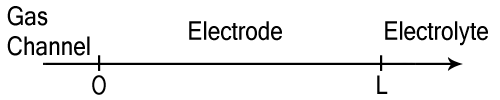}
  \caption{\small Domain schematic for a one-dimensional cathode electrode.} 
  \label{model_domain}
  \end{center}
\end{figure}

The diffusion of gases across the cathode of a MCFC are now studied 
mathematically using three different models.  The first model
considers only diffusion, while the second and third models consider
two different types of diffusion as well as convection.

The electro-chemical reaction in the cathode,
Equation~(\ref{cathode_react}), involves a reaction between three
constituents and the production of another.  The reactants diffuse
across the cathode from the channel to the elec\-trode/elec\-trolyte
boundary (left to right across the domain shown in
Figure~\ref{model_domain}), while the ions also move from left to
right.  Therefore, a system of four equations will be used that
describe the flow of gases, electrons, and ions. 

\subsection{Fickian Diffusion}
\label{fick}

The first model is derived using Fick's Laws of Diffusion and Ohm's
Law.

The solid and liquid potentials are given by the change in current
density using Ohm's Law and the Butler-Volmer equation 
\begin{align}
  \frac{d}{dx} \left(\sigma_s\epsilon_s(x)^b\frac{d\phi_s}{dx}\right)
  &=-\nu_sS(\phi_s,\phi_l,c_{\ms O_2},c_{\ms CO_2}),
  \label{phi_s}\\
  \frac{d}{dx}\left(\sigma_l\epsilon_l(x)^b\frac{d\phi_l}{dx}\right)
  &=\nu_lS(\phi_s,\phi_l,c_{\ms O_2},c_{\ms CO_2}), 
  \label{phi_l}
\end{align}
where $\sigma_l$ and $\sigma_s$ are the conductivities for the liquid
and solid phase, respectively.  The solid ($\sf{e^-}$) conductivity in
the MCFC is generally one to two orders of magnitude larger than the
liquid ($\sf{CO_3^=}$) conductivity, which will create a near constant
potential for the former across the domain.  The potential is
described by $\phi_l$ and $\phi_s$, and determined by the porosity
$\epsilon_l(x)$ and $\epsilon_s(x)$, stoichiometric coefficients $\nu_s$ and
$\nu_l$, and the Butler-Volmer equation, $S$, describing the reaction
kinetics.   

Using Fick's Second Law, the rate of change of concentration as a
result of diffusion by $\sf{O_2}$ and $\sf{CO_2}$ and their reaction
is found to be 
\begin{align}
  \frac{d}{dx}\left(-D\epsilon_g(x)^b\frac{dc_{\ms O_2}}{dx}\right)
  &=-\nu_{\ms O_2}S(\phi_s,\phi_l,c_{\ms O_2},c_{\ms CO_2}), 
  \label{c_o} \\
  \frac{d}{dx}\left(-D\epsilon_g(x)^b\frac{dc_{\ms CO_2}}{dx}\right)
  &=-\nu_{\ms CO_2}S(\phi_s,\phi_l,c_{\ms O_2},c_{\ms CO_2}), 
  \label{c_co}
\end{align}
where $D$ represents the diffusion coefficient, which is the same value for
both gas species.  Note the exponential dependency on the porosity
$\epsilon_g(x)$.  The constant $b$ represents the Bruggeman
correction for the impedance of diffusion in each phase~\cite{bruggeman:35a}.

The Butler-Volmer equation is given by,
\begin{align}
  S(\phi_s,\phi_l,c_{\ms O_2},c_{\ms CO_2}) &= i_0c_{\ms O_2}c_{\ms
  CO_2}e^{\frac{\eta\alpha nF}{RT}}, &\eta=\phi_s-\phi_l,
  \label{BV_eqn_1}
\end{align}
where $\eta$ represents the difference between solid and liquid potentials,
$n$ represents the number of electrons in the reaction, and $i_0$ the exchange
current density.  The negative exponential term in~(\ref{BV_eqn_1}) is
dropped since the polarization is at least $\eta>0.05\,\rm{V}$.   

The correction factor for the diffusivities and conductivities are
given by the porosity, $\epsilon_i$.  The volume fractions must add up
to one.  If the porosity of two quantities is known, the third
quantity is given by, 
\begin{equation}
  \epsilon_l(x)+\epsilon_g(x)+\epsilon_s(x)=1.
\end{equation}
Equations~(\ref{phi_s})-(\ref{c_co}) can be solved numerically using
Newton's Method for computational efficiency or iterative methods.  

\subsection{Fickian Convection-Diffusion}
\label{fickconv}

Using a convection-diffusion model, the flux becomes the sum
of the convective flux and the diffusive flux.  In this model, the
molecular interactions are not considered explicitly. They will be
included in the third model (Section~\ref{binary}). 

The convection-diffusion equations for $\sf{O_2}$ and $\sf{CO_2}$ are
given by
\begin{align}
  \frac{d}{dx}\left(uc_{\ms O_2}-D\epsilon_g(x)^b\frac{dc_{\ms
  O_2}}{dx}\right) &=-\nu_{\ms O_2}S(\phi_s,\phi_l,c_{\ms
  O_2},c_{\ms CO_2}), \label{c_o_fc} \\
  \frac{d}{dx}\left(uc_{\ms CO_2}-D\epsilon_g(x)^b\frac{dc_{\ms
  CO_2}}{dx}\right) &=-\nu_{\ms CO_2}S(\phi_s,\phi_l,c_{\ms
  O_2},c_{\ms CO_2}), \label{c_co_fc}
\end{align}
where the fluid flux, $u$, is described by Darcy's Law for porous
media,
\begin{equation}
u=-\frac{\kappa\epsilon(x)}{\mu}\frac{dP}{dx}=-\frac{\kappa\epsilon(x)}{\mu}RT\frac{dc_{\ms T}}{dx}.
\label{darcyslaw}
\end{equation}
Compared to the Fickian Diffusion model, there is an extra
term and Newton's method is used to solve this type of non-linear
problem.  Note that the two diffusive fluxes do not necessarily add up
to zero. 

\subsection{Multi\-component Convection-Diffusion}
\label{binary}

The Maxwell-Stefan equations are used to
describe the flux of the diffusing species as well as the interactions
between different molecules in the fluid flow.  The flux for this
model is the sum of the convective flux and the multi\-component
diffusive fluxes.  The conservation equations for the concentration
are given by 
\begin{align}
  \frac{d}{dx}\left(uc_{\ms O_2}-D\epsilon_g(x)^bc_{\ms
  T}\frac{d}{dx}\frac{c_{\ms O_2}}{c_{\ms T}}\right) &=-\nu_{\ms
  O_2}S(\phi_s,\phi_l,c_{\ms O_2},c_{\ms CO_2}), \label{c_o_bc} \\ 
  \frac{d}{dx}\left(uc_{\ms CO_2}-D\epsilon_g(x)^bc_{\ms
  T}\frac{d}{dx}\frac{c_{\ms CO_2}}{c_{\ms T}}\right) &=-\nu_{\ms
  CO_2}S(\phi_s,\phi_l,c_{\ms O_2},c_{\ms CO_2}). \label{c_co_bc} 
\end{align}
This model is similar to the model presented in White {\em et al}.
\cite{white:2003b}, with the addition of convection.

These equations become highly non-linear due to the total
concentration term, $c_{\ms T}$, which also appears in Darcy's law
($c_{\ms T}=c_{\ms O_2}+c_{\ms CO_2}$). Newton's Method can be used to
solve this model but it is more computationally expensive than
the Fickian convection-diffusion model. Note that here the two
diffusive fluxes always add up to zero. 

\subsection{Boundary Conditions}
\label{bcs}

The gas enters the electrode at the channel while the liquid
electrolyte penetrates the electrode pores at the opposite side.  The
gas is unable to flow into the electrolyte due to the pore filling and
the electrode is manufactured in such a way as to avoid the corrosive
electrolyte penetration of the gas channels.

\subsubsection{At the Channel $(x=0)$}
At the channel, the gas is flowing into the electrode.  This boundary
uses Dirichlet conditions that give the value for the concentration
of $\sf{O_2}$ and $\sf{CO_2}$ as well as the solid potential.  The electrolyte
is not allowed to move from the electrode into the channel and zero
flux is enforced using a Neumann condition for the liquid potential.
The boundary conditions are chosen as
\begin{align}
  \phi_s(0)&=\phi_{s,0},&
  \frac{d\phi_l}{dx}(0)&=0,&
  c_{\ms O_2}(0)&=c_{{\ms O_2},0},&
  c_{\ms CO_2}(0)&=c_{{\ms CO_2},0}.
\end{align}
In reality, the electrolyte distribution in the cathode is non-uniform
and $\epsilon_l(x)$ will be zero for a finite range $0\leq x \leq
x_0$.  In fact, $\epsilon_l(x)$ is usually a monotonically increasing
function.  In essence, we are moving the channel from $x=0$ to $x=x_0$
in this work since there are no reactions between $0$ and $x_0$.

\subsubsection{At the Cathode/Electrolyte Interface $(x=L)$}
At the cathode/electrolyte interface, the gas and electrons are not
allowed to enter the electrolyte and this is reinforced using Neumann
conditions.  The liquid potential is given by a Dirichlet condition.
These boundary conditions are chosen to be
\begin{align}
  \frac{d\phi_s}{dx}(L)&=0,&
  \phi_l(L)&=\phi_{l,L},&
  \frac{dc_{\ms O_2}}{dx}(L)&=0,&
  \frac{dc_{\ms CO_2}}{dx}(L)&=0.
\end{align}
\begin{figure}[t]
 \begin{minipage}[t]{0.47\linewidth}
   \includegraphics[width=\linewidth]{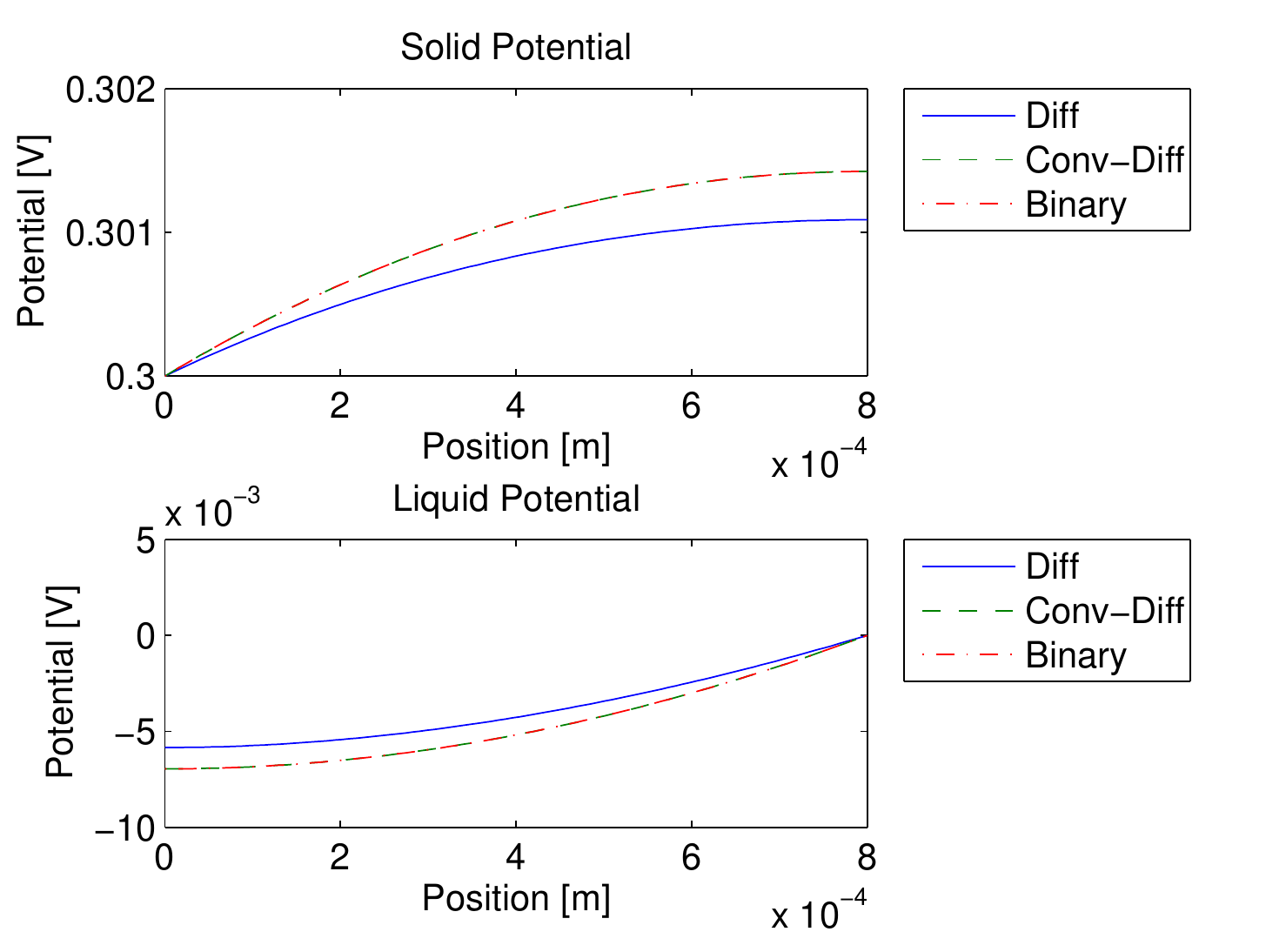}
   \caption{\small The solid and liquid potential distribution across the
   electrode using the reference parameters as stated in the
   nomenclature.  The Fickian convection-diffusion (Conv-Diff) and
   multi-component convection-diffusion (Binary) models share the same
   profile.}
   \label{efig1}
 \end{minipage}
 \hspace{0.5cm}
 \begin{minipage}[t]{0.47\linewidth}
 \includegraphics[width=\linewidth]{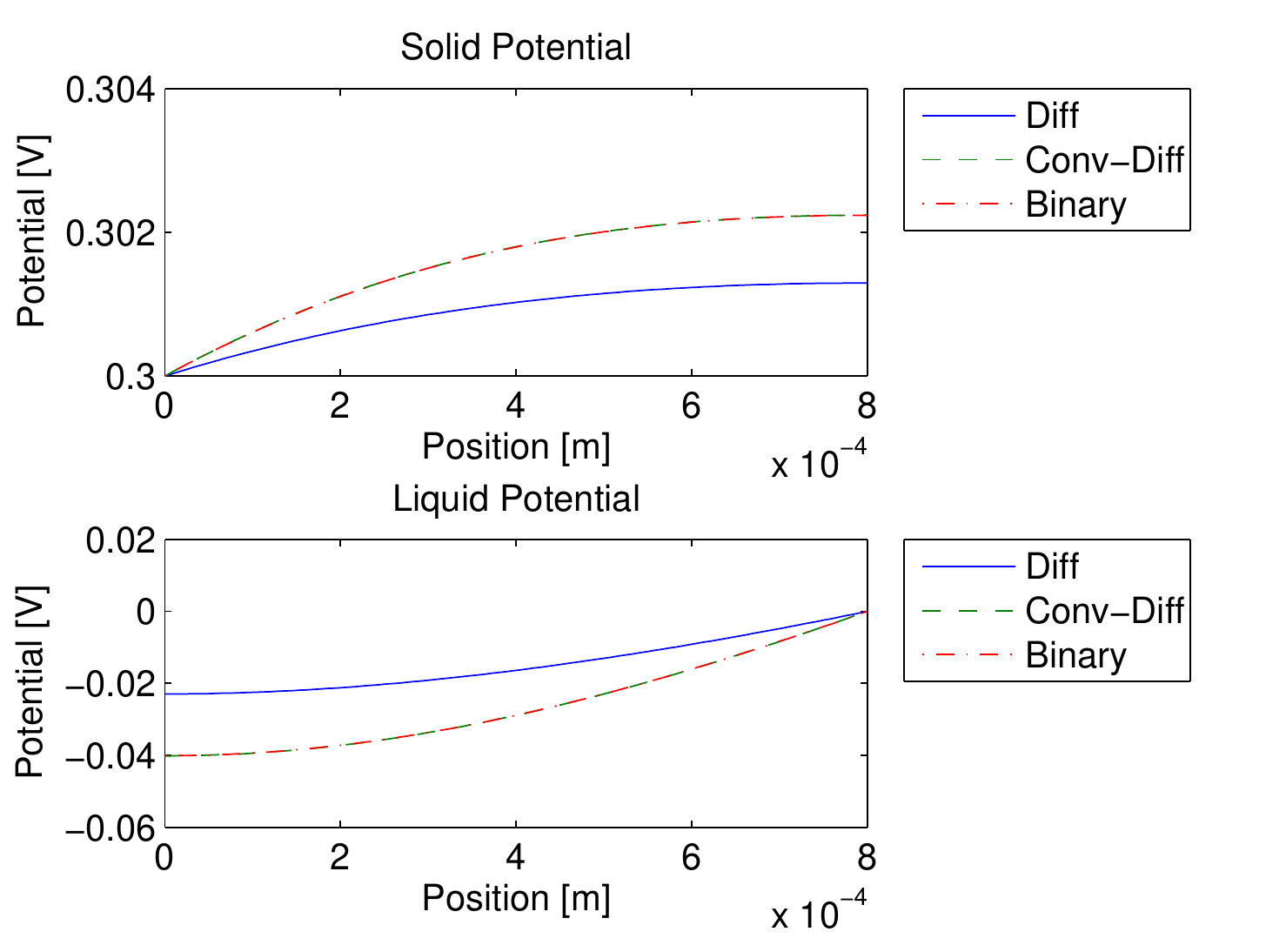}
 \caption{\small The potential distribution with the liquid conductivity
   decreased from the reference value of $140\,\rm{S/m}$ to $50\,\rm{S/m}$, which
   increases the reaction rate.}
 \label{efig2}
 \end{minipage}
\end{figure}

\begin{figure}[t]
 \begin{minipage}[t]{0.47\linewidth}
   \includegraphics[width=\linewidth]{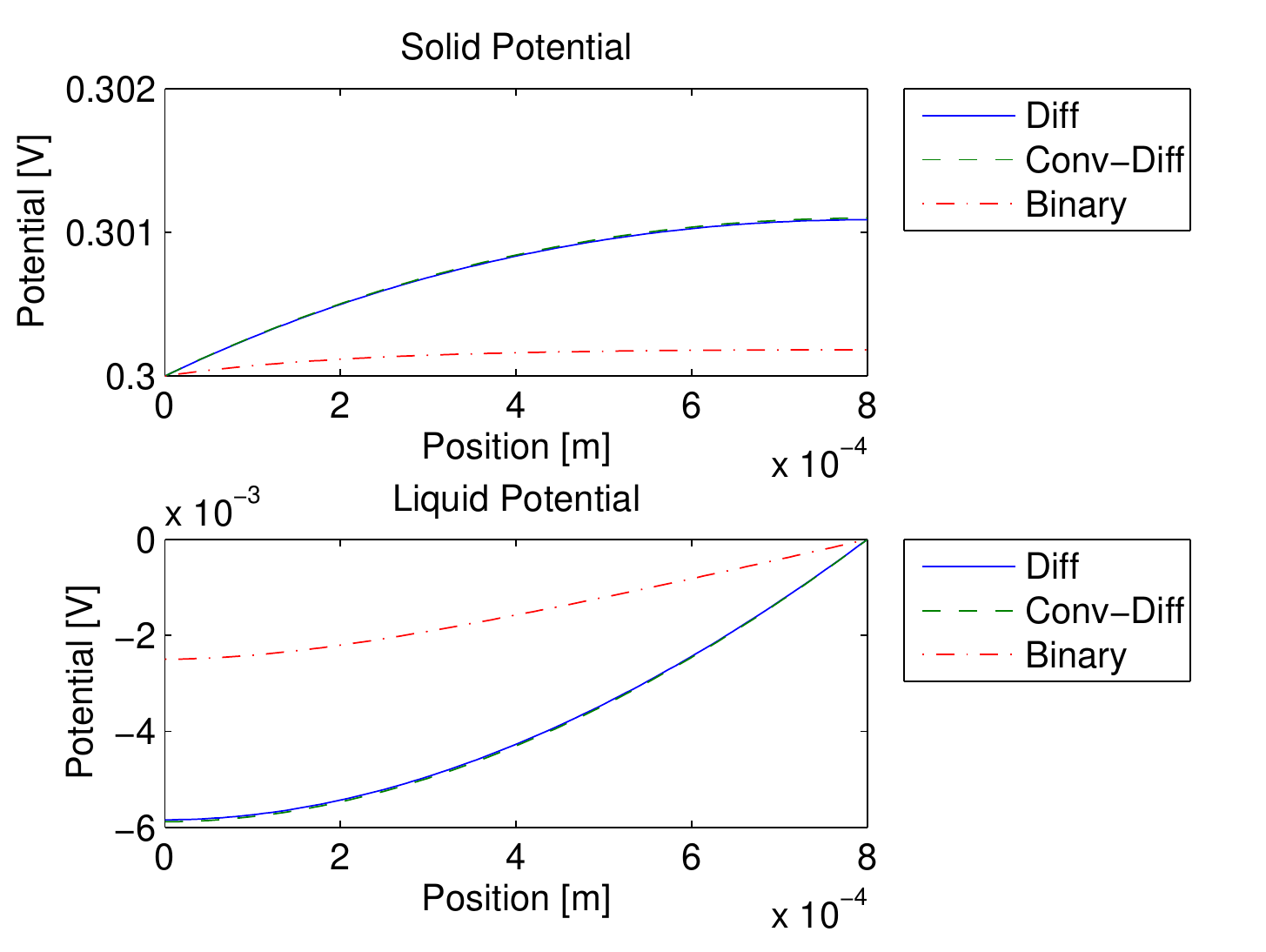}
   \caption{\small The potential distribution with the permeability decreased
   three orders of magnitude effectively turning off the convective
   flux.}
   \label{efig3}
 \end{minipage}
 \hspace{0.5cm}
 \begin{minipage}[t]{0.47\linewidth}
   \includegraphics[width=\linewidth]{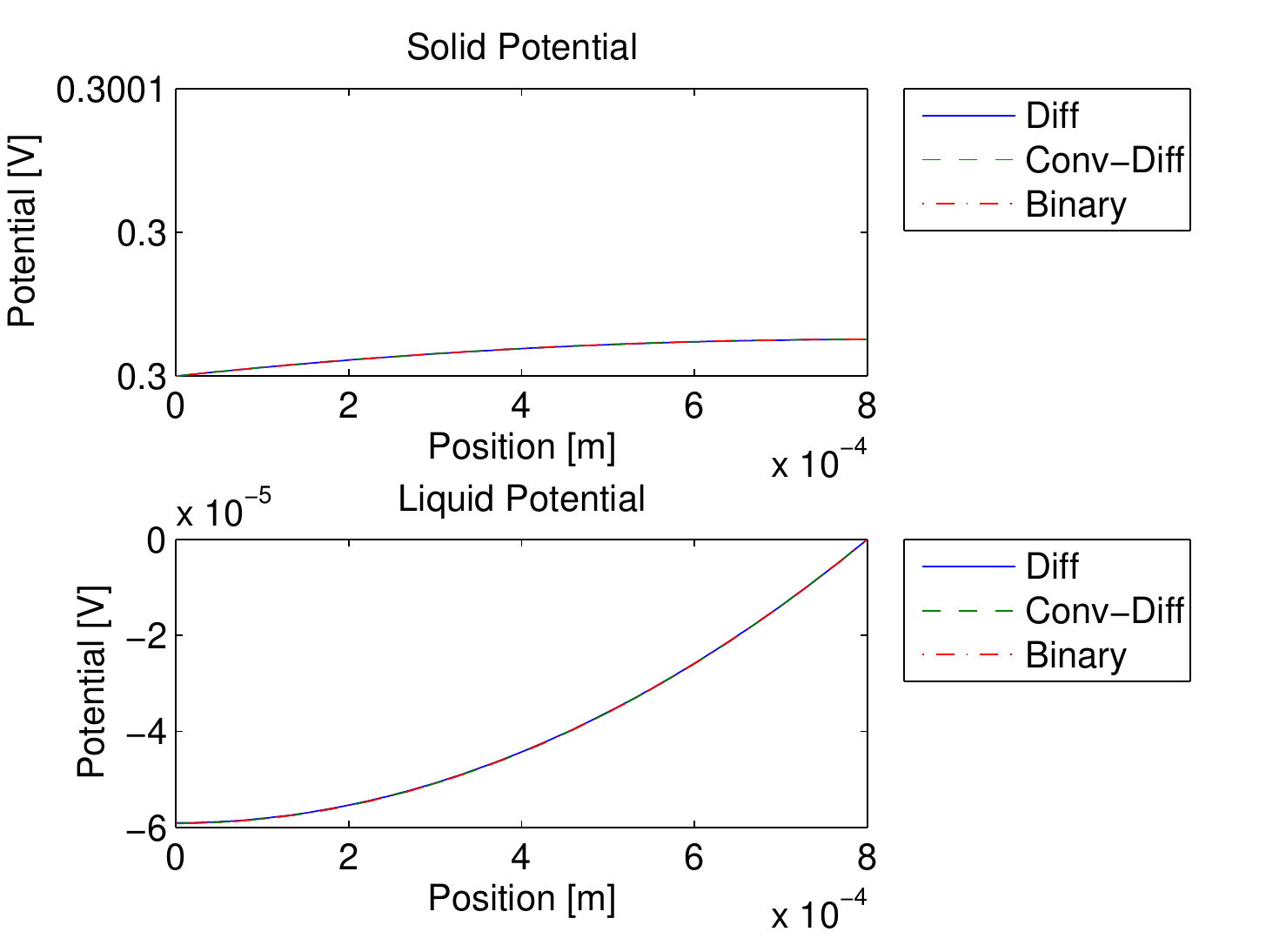}
   \caption{\small The potential distribution with the exchange current
     density decreased by two orders of magnitude, slowing the reaction
     rate.  The solid potential varies less than $10^{-3}\,\rm{V}$
     across the domain.}
   \label{efig4}
 \end{minipage}
\end{figure}

\subsection{Results}

\subsubsection{Solid and Liquid Potential}
The potential drop is shown by plotting the solid and liquid phase
potentials in Figures \ref{efig1}-\ref{efig4}.

The potential difference across the domain using the reference values
as in the nomenclature is plotted in Figure \ref{efig1}.  The
potential remains relatively constant across the domain and all three
models remain within one-thousandth of a decimal point in agreement.  

The steady state solution for the potential difference, as the liquid
conductivity is decreased, is plotted in Figure \ref{efig2}.  The liquid
conductivity, $\sigma_l$, is decreased from $140\,\rm{S/m}$ to
$50\,\rm{S/m}$.  If the conductivity is decreased further, the steady
state solution ultimately does not exist and this will be examined in
Section~\ref{cond_existence}.  As the liquid conductivity is
decreased, the potential gradient increases, hence, the reaction rate
increases and since all four equations are coupled, the solid
potential and species concentrations also change.  This increase
is attributed to the inverse relationship between the rate of change
in liquid potential and conductivity.  The convection-diffusion and
multi\-component models share the same profile despite the differences
in the formulation.  All three models have a change within the same
order of magnitude. 

The value for the permeability of the MCFC cathode is difficult to
obtain but is comparable to that of a Proton Exchange Membrane fuel
cell (PEMFC) gas diffusion layer, which contains a similar pore size.
The potential difference, when the permeability, $\kappa$, is decreased
by three orders of magnitude, causing the convective flux to approach
zero, is plotted in Figure \ref{efig3}.  With the convective flux near
zero, the Fickian diffusion and convection-diffusion models are
essentially identical and this is shown in the graph as they now share
the same profile.  The enforcement of zero net diffusive flux in the
multi\-component model keeps the potential change lower than that
shown by the other models due to the small mass transfer of
gas. If the permeability increases, the convective term dominates
further and will result in similar results as before. 

The exchange current density, $i_0$, enters the Butler-Volmer
equation and can be found to vary by several orders of magnitude
depending on the model. The potential difference is
plotted in Figure \ref{efig4} with the exchange current density at a
value of two orders of magnitude less than the reference
value, $1.0\times10^{-3}\,\rm{A/m^2}$, which is the value used in
White {\em et al}.~\cite{white:2003b}.  By decreasing the exchange
current density, the reaction rate for each  species decreases and the
rate of change decreases as well.  At such low reaction rates, each
model shares the same profile across the domain.

\subsubsection{Concentration}

\begin{figure}[t]
  \begin{minipage}[t]{0.47\linewidth}
    \includegraphics[width=\linewidth]{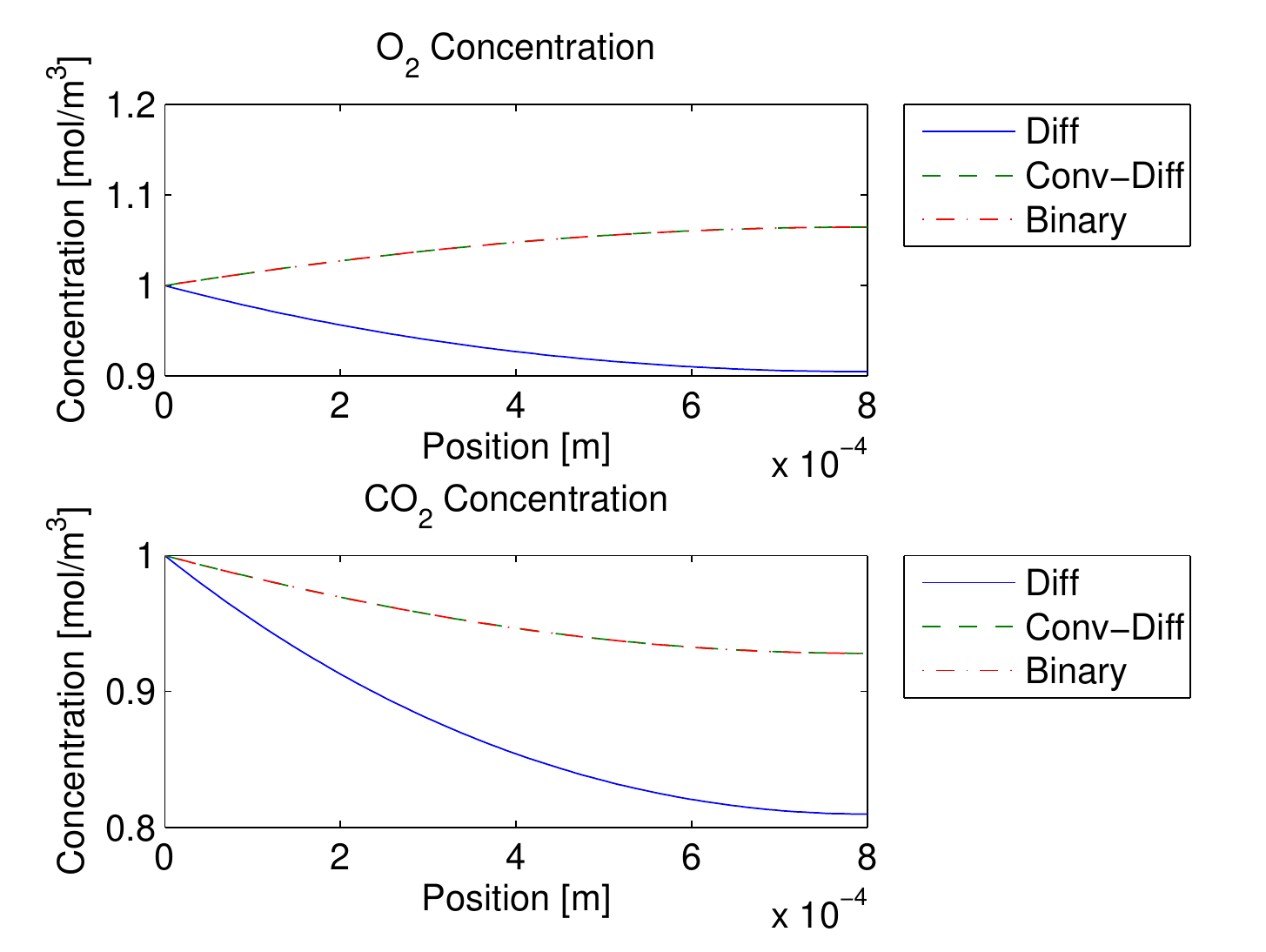}
    \caption{\small The $\sf{O_2}$ and $\sf{CO_2}$ concentration
      profiles across the electrode using the reference values. The
      Fickian convection-diffusion (Conv-Diff) and multi-component
      convection-diffusion (Binary) models share the same profile.}
      \label{cfig1}
  \end{minipage}
  \hspace{0.5cm}
  \begin{minipage}[t]{0.47\linewidth}
    \includegraphics[width=\linewidth]{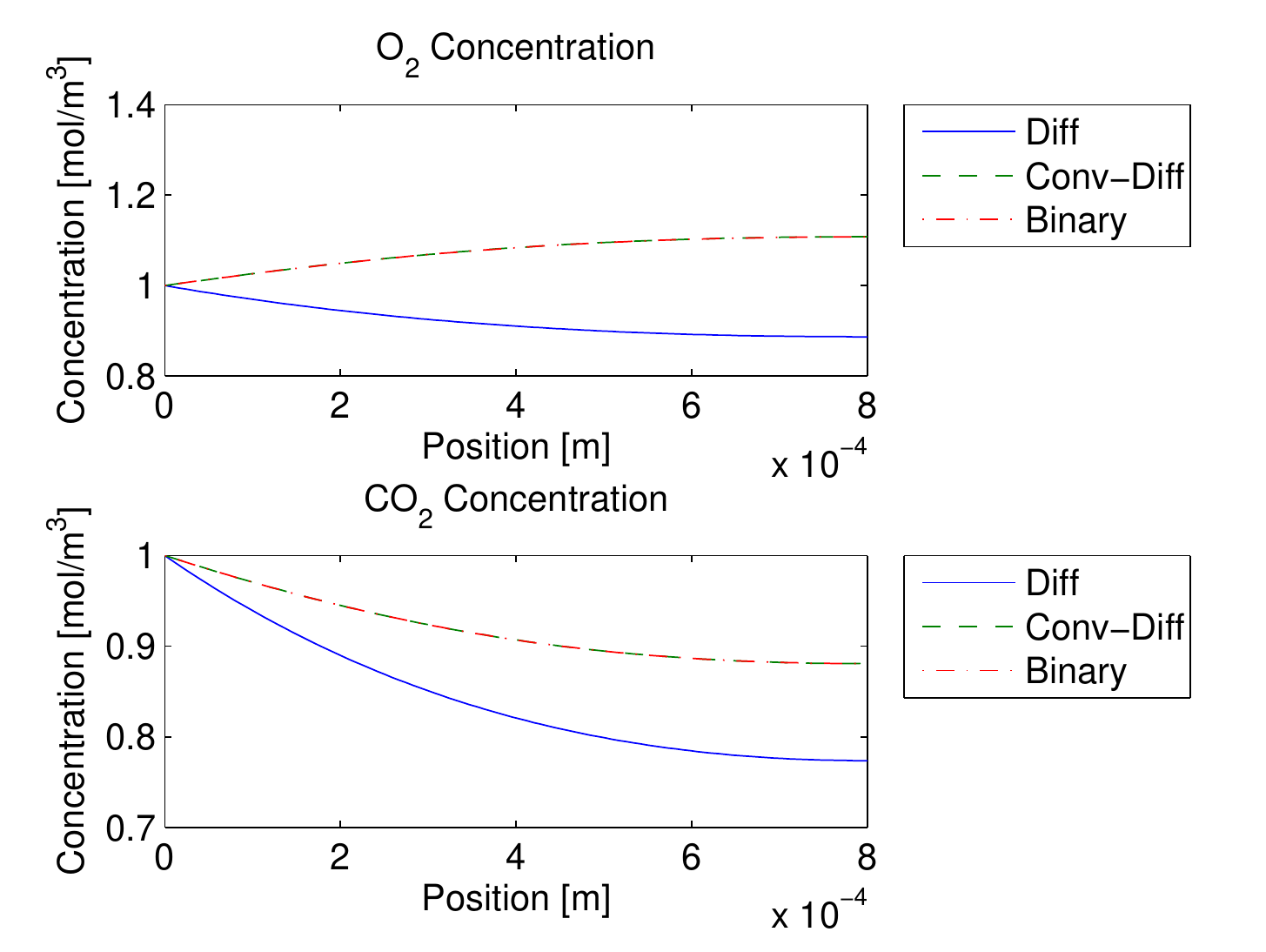}
    \caption{\small The concentration profiles with the liquid conductivity
      decreased from the reference value of $140\,\rm{S/m}$ to $50\,\rm{S/m}$, which increases the
      reaction rate for the liquid potential.}
    \label{cfig2}
  \end{minipage}
\end{figure}

\begin{figure}[t]
  \begin{minipage}[t]{0.47\linewidth}
    \includegraphics[width=\linewidth]{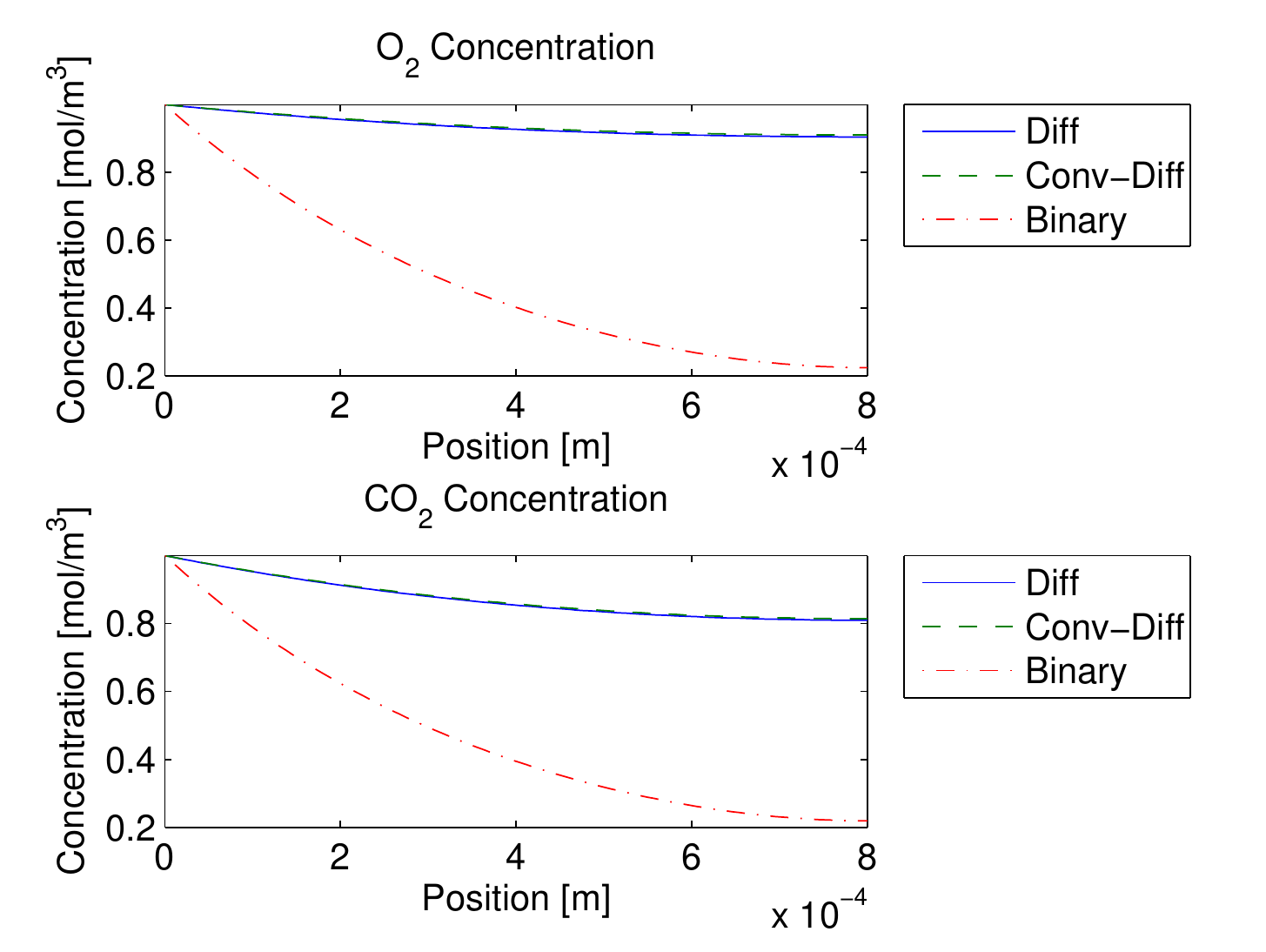}
    \caption{\small The concentration profiles with the permeability
      decreased three orders of magnitude, effectively turning off the
      convective flux.}
    \label{cfig3}
  \end{minipage}
  \hspace{0.5cm}
  \begin{minipage}[t]{0.47\linewidth}
    \includegraphics[width=\linewidth]{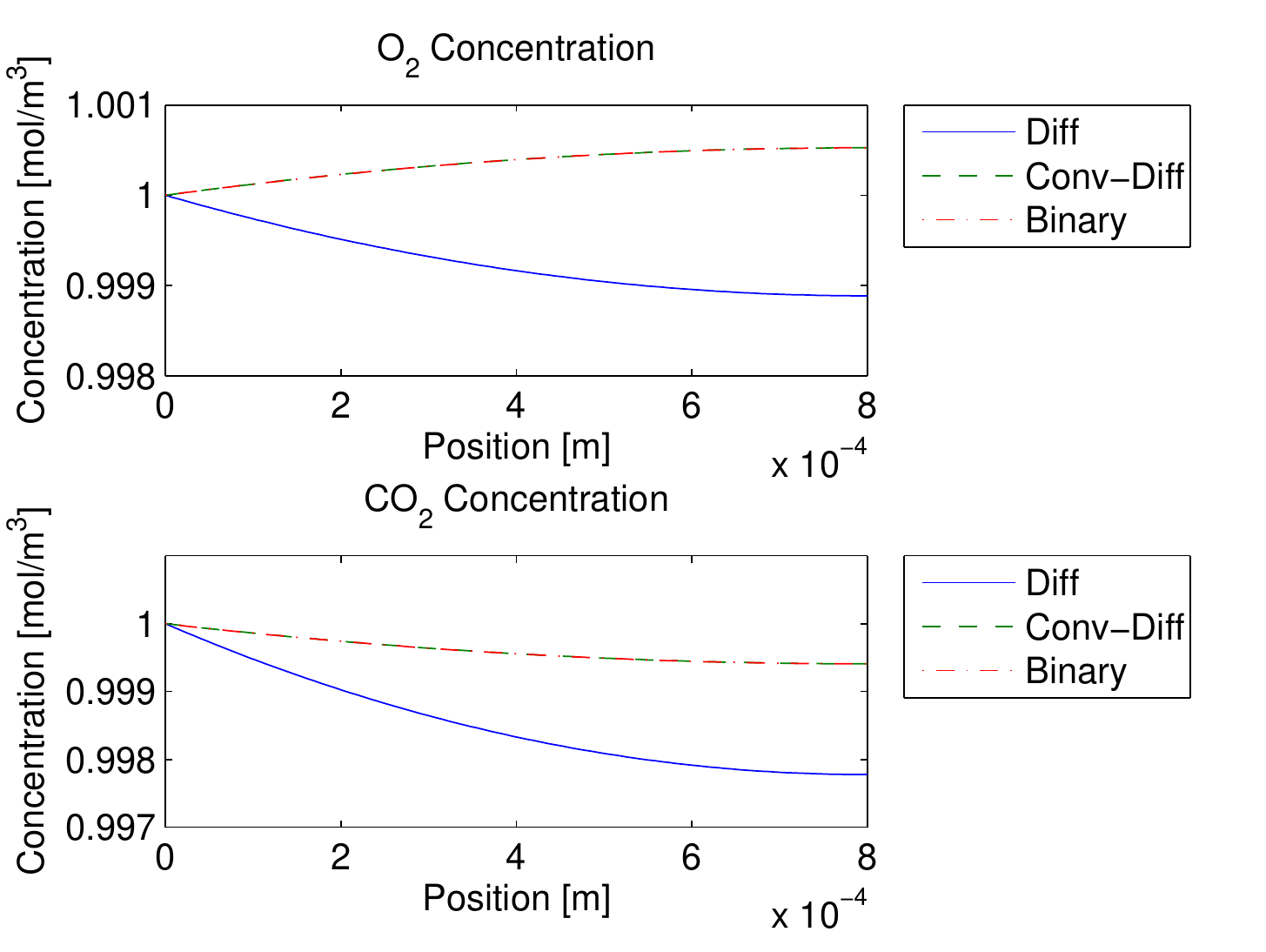}
    \caption{\small The concentration profiles with the exchange current
      density decreased by two orders of magnitude, slowing the reaction rate.} 
    \label{cfig4}
  \end{minipage}
\end{figure}

The change in concentration of $\sf{O_2}$ and $\sf{CO_2}$ is plotted in Figures
\ref{cfig1}-\ref{cfig4}, corresponding to the results for the potential
difference.

The convection term dominates the diffusion term for $\sf{O_2}$ across the
electrode as it pushes the gas towards the electrolyte as in Figure
\ref{cfig1}, using the reference values in the nomenclature.  This can
be seen due to the increase in $\sf{O_2}$ concentration as it
approaches the cathode/electrolyte boundary.  With convection
dominating, there is less than a 10\% change in concentration, while
the Fickian diffusion model without convection shows a 10-20\% change
in $\sf{O_2}$ and $\sf{CO_2}$ concentration.  

Considering the concentration of $\sf{O_2}$, it increases across the domain due
to convection and it is not possible to approximate the
convection-diffusion profile using an effective diffusivity for the
Fickian diffusion model as in White {\em et al}.~\cite{white:2003b}, which
did not include convection.

The steady state solution for a decrease in the liquid conductivity,
$\sigma_l$, is shown in Figure \ref{cfig2}.  The value is decreased by
an order of magnitude and if decreased further, the steady state
solution does not exist (see Section~\ref{cond_existence}) and the
code no longer converges.  Since the system of four differential
equations is coupled, the gradient of the concentration of the
gas-species changes along with the changes in reaction rate.  The
convective flux remains dominant and the convection-diffusion and
multi\-component models continue to share the same profile.

The concentration profile is shown in Figure \ref{cfig3} when the
permeability, $\kappa$, is decreased by three orders of magnitude.
Since the convection is close to zero, diffusion dominates and the
Fickian diffusion and convection-diffusion models are essentially the
same as was described previously for the potential difference.  The
differences between the convection-diffusion and multi\-component models
becomes more apparent as the concentration drops almost 80\% across
the cathode using the latter model, while less than 20\% for the
former model.  This can be attributed to maintaining a non-zero
convective flux.  The discrepancy between the two models will be
further analyzed in Section~\ref{conv_comparison}.

The exchange current density, $i_0$, when decreased by two orders of
magnitude, decreases the reaction rates for all species.  As the
reaction rate decreases, less species react at the three-phase
boundary and the concentration drop is minimal across the domain.
With the permeability at the standard value, convection dominates even
with low reaction rates.  Once again, this is the value of the
exchange current density reported in White {\em et al}.~\cite{white:2003b}.
It still shows the convection term dominating which must be considered
in the mass transport of the MCFC cathode. 

It should be pointed out here that the model by White {\em et al}.
\cite{white:2003b} is ill-posed in that it neglects convection while
keeping binary-diffusion terms.  Since the latter add up to zero, we
cannot have a net flux of gas across the electrode, an obvious
contradiction, which they do not discuss.  This can be seen by adding
(\ref{c_o_bc}) and~(\ref{c_co_bc}) giving (for $u=0$), 
\begin{align}
  \frac{d}{dx}\left(-D\epsilon_g(x)^bc_{\ms T}\frac{d}{dx}
  \left(\frac{c_{\ms O_2}}{c_{\ms T}} + \frac{c_{\ms CO_2}}{c_{\ms
      T}}\right)\right) &= -(\nu_{\ms O_2}+\nu_{\ms
    CO_2})S(\phi_s,\phi_l,c_{\ms O_2},c_{\ms CO_2}),\nonumber \\
  &=\frac{d}{dx}\left(-D\epsilon_g(x)^bc_{\ms T}\frac{d}{dx}(1)\right)
  = 0,
\end{align}
which is a contradiction since $S$ is not identically zero.

\subsubsection{Cell Performance}

\begin{figure}[t]
 \centering
 \includegraphics[width=0.8\linewidth]{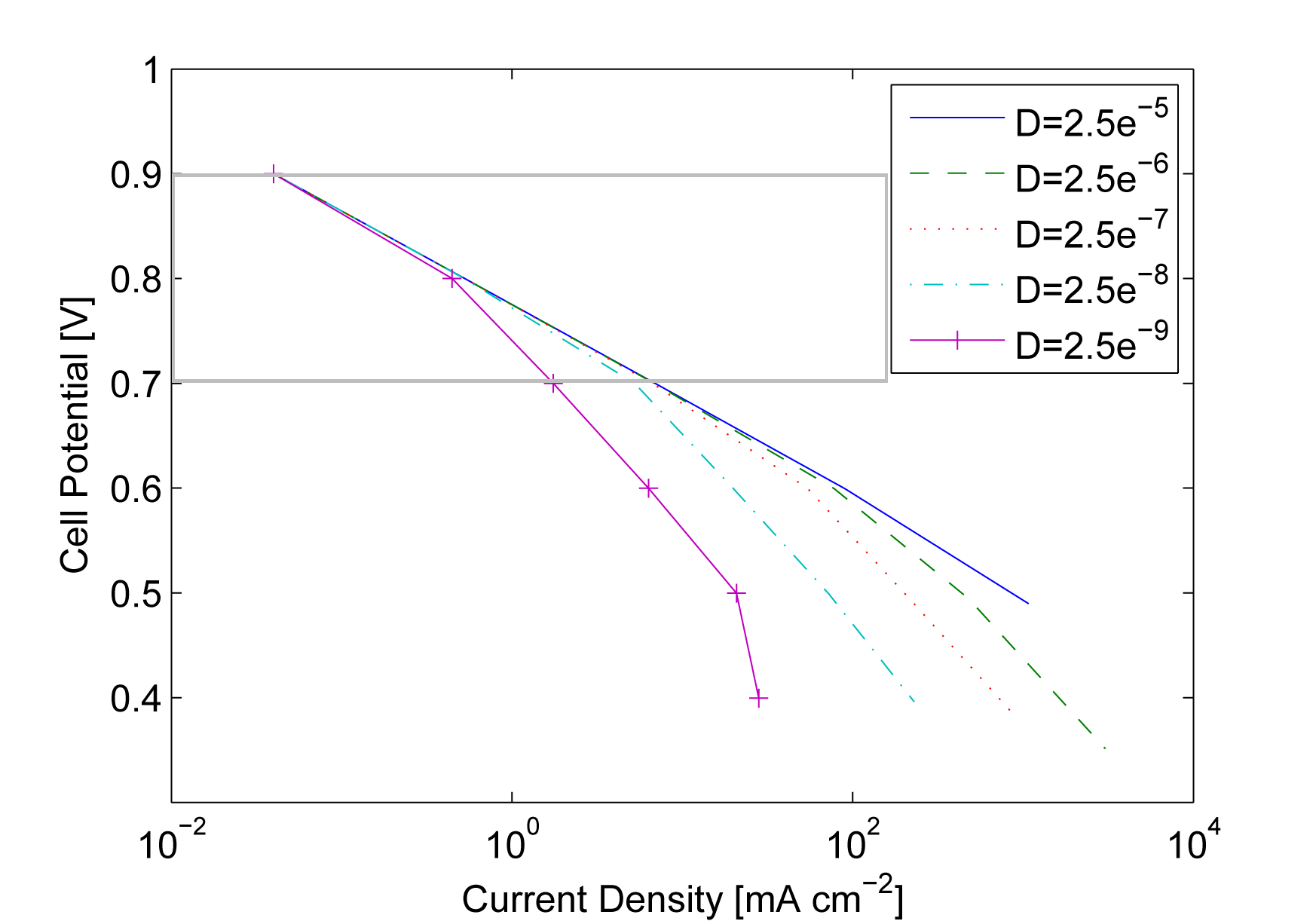}
 \caption{\small Cell potential versus current density: polarization curve for
   the half cell Fickian convection-diffusion model.} 
 \label{performance}
\end{figure}

The MCFC operates at current densities in the range of
$100-200\,\rm{mA/cm}^2$ with a cell potential of $0.75-0.90\,\rm{V}$.
Figure \ref{performance} shows the polarization curve for the
convection-diffusion model at several different values of the
diffusivity, $D$.  As the diffusivity decreases, the polarization
curve loses its linear character as the degree of irreversibility
increases.  At the standard reference value for the diffusivity,
$2.5\times10^{-5}\,\rm{m^2/s}$, the cell operates in the appropriate
range for the current density and potential difference.  This graph
was made using an open-circuit or reversible cell potential of
$E_r=1.0\,\rm{V}$, which is appropriate for MCFCs \cite{white:2003b}.
The expected cell polarization for the half-cell reaction in the
cathode is shown in the figure as a box. 

\subsubsection{Comparison of Convection-Diffusion Models}

\label{conv_comparison}

In many cases, researchers in the field of porous media flow are
interested in approximating the convection-diffusion equation by using
only Fickian diffusion with an effective diffusivity that will provide
a similar profile with convection excluded.  In the case when the
convection term is included along with Fickian diffusion, the
concentration of $\sf{O_2}$ can increase from the channel to the
electrolyte when convection is in the opposite direction of diffusion.
Due to this increase, it is not possible to use an effective
diffusivity that will provide the same profile since diffusion will
only cause the concentration to decrease across the domain.  This
means that convection must be included in the mass transport of the
MCFC cathode and can not be neglected as in White {\em et al}.~\cite{white:2003b}. 

Another area of interest in the study of mass transport in fuel cells is
the comparison between the convection-diffusion model of
Section~\ref{fickconv} and the multi\-component convection-diffusion
model of Section~\ref{binary}.  Depending upon the parameters, it is
possible that Fick's Law can provide a sufficient description of
diffusion across a domain.  Note that in the first model, the
diffusive fluxes do not add up to zero whereas in the second model
they do. 

The permeability, $\kappa$, is a measure of the ability of a fluid to
move through a porous medium.  It can be very difficult to measure the
permeability of a material for fuel cells as it is directly related to
the manufacturing of the material.  Values for permeability in the
literature for MCFCs have not been found but we can safely use the
values for the permeability in the proton exchange membrane fuel cells
(PEMFC), which is the primary fuel cell in use for mobile
applications. Currently, there exists only 2-3 manufacturers of
MCFCs and values for the permeability are hard to come by. The pore
size in the cathode of the PEMFC is approximately
$10\,\rm{\mu}\rm{m}$, which is close to that of the
MCFC. Realistically, we can determine the permeability of the MCFC to
be close to $1.9\times10^{-12}\,\rm{m^2}$ as in the PEMFC model of
Promislow {\em et al}.~\cite{promislow}.    

\begin{figure}[t]
 \centering
 \includegraphics[width=0.8\linewidth]{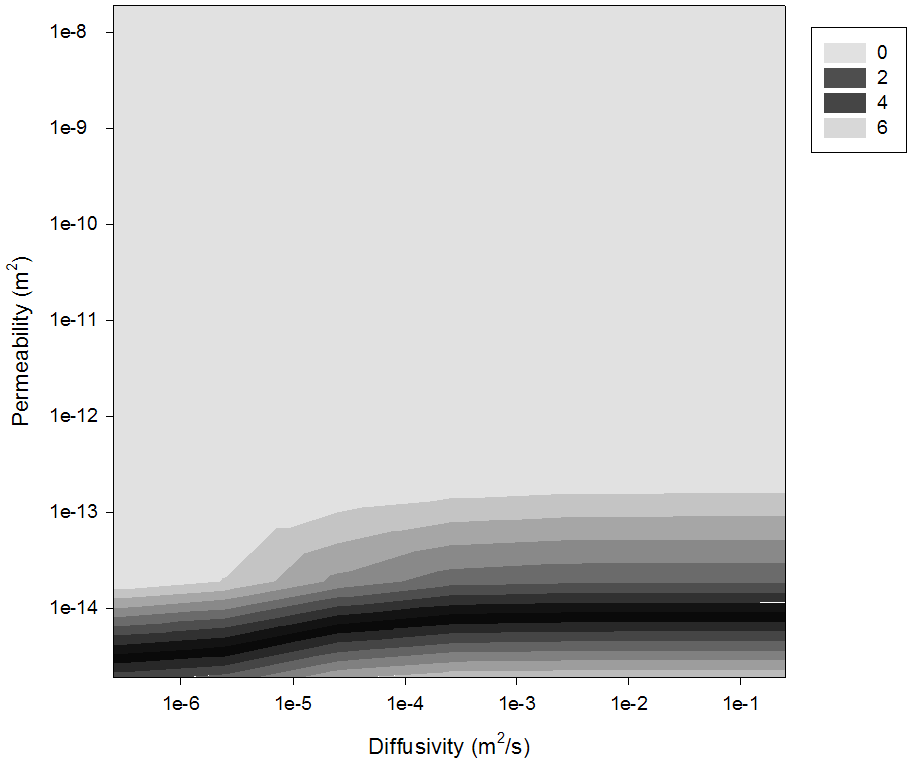}
 \caption{\small The difference between the concentration profiles for both
   convection-diffusion and multi\-component transport models.  For the
   standard values of both diffusivity, $D$, and permeability,
   $\kappa$, the error between the solutions is minimal.  The error
   increases for smaller values of the permeability where the
   convection term is closer to zero.  The colours represent the error
   between profiles and is less than $10^{-5}$ up to permeabilities
   less than $1\times10^{-13}$, where the error grows above 1.} 
 \label{convdiff_stefmax}
\end{figure}

The diffusion coefficients in the convection-diffusion model are
the same for any number of species considered using Fick's First Law
of Diffusion.  In the Stefan-Maxwell formulation, only $n-1$ fluxes
are independent and for $n>2$, the diffusion coefficients consist of a
matrix.  The diffusion coefficients for the species in the cathode of
the MCFC are also difficult to find from literature.  For instance,
White {\em et al}.~\cite{white:2003b} state the diffusion coefficients to
be $1.16\times10^{-4}\,\rm{m^2/s}$, which is referenced from Cussler
\cite{cussler,white:2003b}, but the diffusion of $\sf{O_2}$
through air, which is the highest possible diffusivity, is only 
$0.25\times10^{-4}\,\rm{m^2/s}$. The value reported in Fehribach {\em
  et al}.~\cite{fehribach:96}, $4.3\times10^{-6}\,\rm{m^2/s}$, is
much more realistic.    

Due to the uncertainty of values for the permeability and
diffusivities, a comparison between the two models for different
values is performed so as to validate the use of the convection-diffusion
equations as an approximation to the Stefan-Maxwell equations.  The
convection-diffusion model is far less computationally expensive than
the more non-linear multi\-component convection-diffusion model, which
will be beneficial for scaling-up the model to higher dimensions. 

Figure \ref{convdiff_stefmax} shows the difference (error) between
the $\sf{O_2}$ concentration profiles of the Fickian convection-diffusion and
multi\-component convection-diffusion models.

The error is calculated as the $\rm{L^2}$-norm of the difference
between the $\sf{O_2}$ concentration profiles of the Fickian
convection-diffusion (fcd) and multi-component convection-diffusion
(mcd) models normalized by the area under the $\sf{O_2}$
convection-diffusion profile, so that  
\begin{equation}
{\rm error}=\frac{\left(\int^L_0 |c_{\ms O_2}^{\ms fcd}-c_{\ms
    CO_2}^{\ms mcd}|^2\,dx\right)^{1/2}}{\int^L_0 c_{\ms O_2}^{\ms
    fcd}\,dx},
\end{equation}
where Simpson's rule was used to find the area.

For permeability values above $10^{-13}\,m^2$, there is a very small
difference between the profiles (less than $10^{-5}$).  For smaller
values of the permeability, the difference increases significantly.
However, for realistic values based on the known diffusivities from
the MCFC literature and permeabilities from the PEMFC literature, the
convection-diffusion model is in very good agreement with the
multi\-component model. 

Based on these results, it seems justified to approximate the
multi\-component model by using the simpler, and computationally less
expensive, convection-diffusion model in MCFC cathode electrodes.
This is beneficial for scaling-up the model to higher dimensions as
well as in stack models where run time increases dramatically.

\section{Fickian Diffusion: Non-Existence of Steady-State Solutions}
\label{cond_existence}
While performing numerical simulations of the models given in
Section~\ref{fick}, the solution did not converge for small values of
the liquid conductivity.  The value for liquid conductivity,
$\sigma_l$, reported in White {\em et al}.~\cite{white:2003b} is around
$2.0\,\rm{S/m}$ and in Fehribach {\em et al}.~\cite{fehribach:96} it is
around $140\,\rm{S/m}$.  The model presented here will only admit
solutions when the conductivity is well above the value from White
{\em et al}.

The question that arises is whether this is a numerical issue or an
ill-posedness of the model.

Using Equation~(\ref{phi_l}),
\begin{equation}
  \frac{d}{dx}\left(\sigma_l\epsilon_l(x)^b\frac{d\phi_l}{dx}\right)=\nu_li_0c_{\ms
    O_2}c_{\ms CO_2}\exp{\left(\frac{n\alpha F(\phi_s-\phi_l)}{RT}\right)},
\end{equation}
we can solve this analytically by
non-dimensionalizing the equation according to
\begin{equation}
\bar{\phi}_{l,\bar{x}\bar{x}}-\delta e^{-\bar{\phi}_l}=0,
\label{phi_l_nondim}
\end{equation}
where
\begin{align}
  \bar{\phi}_l&=\frac{n\alpha F}{RT}\phi_l,&
  \bar{x}&=L^{-1}x,&
  \delta&=\frac{\nu_{\ms
      CO_3^=}L^2}{\sigma_l\epsilon_l(x)^b}\frac{RT}{n \alpha F} i_0 c_{\ms
    O_2}c_{\ms CO_2}e^{\bar{\phi}_s}. \label{delta_eq}
\end{align}
In this non-dimensionalized model, the porosity $\epsilon_l(x)$ and
solid potential $\phi_s$ are considered to be constant across the
domain, which is a valid assumption based on the results of
Chapter~\ref{models}.  The concentration of $\sf{O_2}$ and $\sf{CO_2}$
are also considered constant across the domain which is a reasonable
first-order approximation.

The parameter $\delta$ contains the liquid conductivity $\sigma_l$.
We will now show that a critical value, $\delta_c$, can be
found beyond which the steady-state solution will not exist. 

\subsection{Thermal Runaway}

The potential equation, Eq.~(\ref{phi_l_nondim}), is similar (replace
$\bar{\phi}_l\to-\bar{\phi}_l$) to the steady-state equation
for thermal runaway found in Fowler~\cite{fowler}
\begin{equation}
\theta_{xx}+\lambda e^{\theta}=0,
\label{fowler_ode}
\end{equation}
with boundary conditions
\begin{align}
\theta(\pm1)&=0.
\end{align}
This system can be solved analytically as follows: multiply both
sides by $\theta_x$ and integrate, yielding
\begin{subequations}
  \begin{align}
    \int \theta_x\theta_{xx} dx&=\int -\lambda e^{\theta} \theta_x dx, \\
    \Rightarrow \frac{1}{2}\theta_x^2&=-\lambda e^{\theta} + C.
  \end{align}
\end{subequations}
The constant $C$ can be found by imposing the symmetry condition
$\theta(-x)=\theta(x)$ and if $\theta$ is smooth enough,
$\theta_x(0)=0$.  We find $C=\lambda e^{\theta_0}$ where
$\theta_0=\theta(0)$.

Considering only the problem on the interval $x=[0,1]$ in comparison
with Eq.~(\ref{phi_l_nondim}), the ODE is re-arranged and the method
of separation of variables is applied which gives
\begin{subequations}
  \begin{align}
    \theta_x&=\sqrt{2\lambda}\sqrt{e^{\theta_0}-e^{\theta}}, \\
    \Rightarrow 
    \int^{\theta}_{\theta_0}
    \frac{d\theta}{\sqrt{e^{\theta_0}-e^{\theta}}}&=\int_0^x
    \sqrt{2\lambda}dx=\sqrt{2\lambda}x.
    \label{sepvars}
  \end{align}
\end{subequations}
Let us set $z=\sqrt{e^{\theta_0}-e^{\theta}}$ so that 
\begin{subequations}
  \begin{align}
    d\theta&=-2ze^{-\theta}dz,\\
    \Rightarrow
    \frac{d\theta}{\sqrt{e^{\theta_0}-e^{\theta}}}&=\frac{-2dz}{e^{\theta_0}-z^2}.
  \end{align}
\end{subequations}
Substitution into (\ref{sepvars}) gives
\begin{align}
  \sqrt{2\lambda}x&=\int_{\theta_0}^{\theta}
  \frac{d\theta}{\sqrt{e^{\theta_0}-e^{\theta}}} = -2\int
  \frac{dz}{e^{\theta_0}-z^2} =
  -2e^{-\theta_0/2}\tanh^{-1}\frac{z}{e^{\theta_0/2}}, 
\end{align}
where we have used the integral~\cite{grosche}
\begin{align}
  \int \frac{du}{a^2-u^2}&=\frac{1}{a}\tanh^{-1}\frac{u}{a}.
\end{align}
Re-arranging for $z$ leads to
\begin{equation}
  z=\sqrt{e^{\theta_0}-e^{\theta}}=e^{\theta_0/2}\tanh\left(-\sqrt{\frac{\lambda}{2}e^{\theta_0}}x\right)=-e^{\theta_0/2}\tanh\left(\sqrt{\frac{\lambda}{2}e^{\theta_0}}x\right),
\end{equation}
using the identity $\tanh(-x)=-\tanh(x)$.  Defining
$\gamma=\sqrt{\frac{\lambda}{2}e^{\theta_0}}x$, we obtain 
\begin{subequations}
  \begin{align}
    e^{\theta}&=e^{\theta_0}\left(1-\tanh^2\gamma\right)=e^{\theta_0}\sech^2\gamma
  \end{align}
  or
  \begin{align}
    \theta=\theta_0-2\ln\cosh\gamma,
    \label{theta_sol}
  \end{align}
\end{subequations}
using the identity
$1-\tanh^2\gamma=\sech^2\gamma=\cosh^{-2}\gamma$. This solution for
$\theta$ was found in Fowler~\cite{fowler} and we can use the same
approach to solve for the liquid potential.

\begin{figure}[t]
  \centering
  \includegraphics[width=0.8\linewidth]{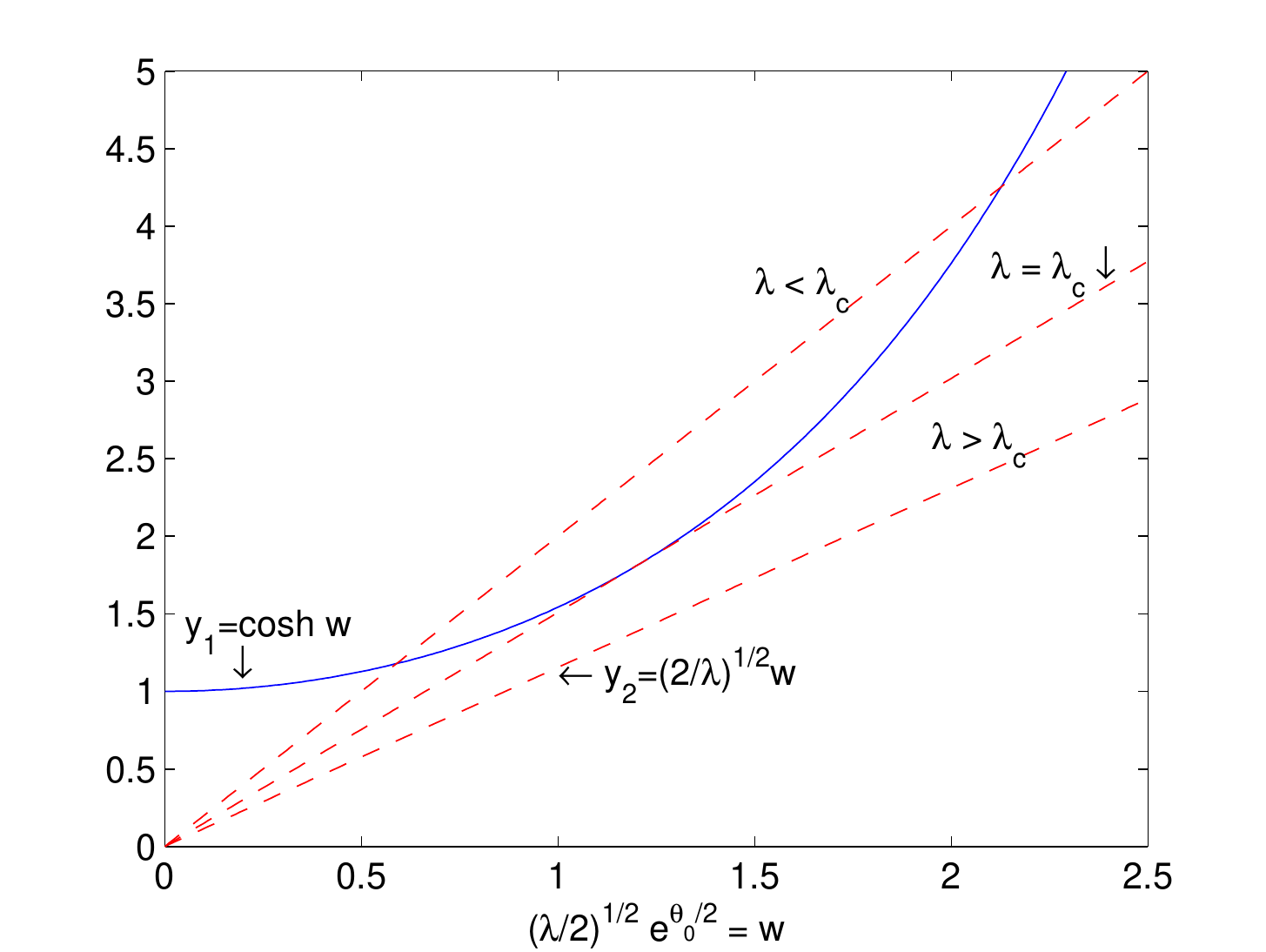}
  \caption{\small In order to obtain solutions to Eq.~(\ref{maxtemp}), the
    possible solutions based on different values of $\lambda$ are shown
    graphically.  As $\lambda$ decreases, the number of solutions
    increase~\cite{fowler}.} 
  \label{solmaxtheta0}
\end{figure}

For this model based on thermal runaway, the maximum temperature,
$\theta_0$, occurs at $x=0$ (by symmetry), and is determined by
satisfying the boundary condition at $x=1$, $\theta(1)=0$, and so from
(\ref{theta_sol}) 
\begin{equation}
  e^{\theta_0/2}=\cosh\left(\sqrt{\frac{\lambda}{2}}e^{\theta_0/2}\right).
  \label{maxtemp}
\end{equation}
The solutions to this transcendental equation are studied in Figure
\ref{solmaxtheta0}, which was re-created based on the figure presented
in Fowler~\cite{fowler}.  Based upon these results, there will either
be 2, 1 or 0 solutions depending on whether $\lambda<\lambda_c$,
$\lambda=\lambda_c$ or $\lambda>\lambda_c$, respectively. The critical value,
$\lambda_c$, is found in the following manner.

\begin{figure}[t]
  \centering
  \includegraphics[width=0.8\linewidth]{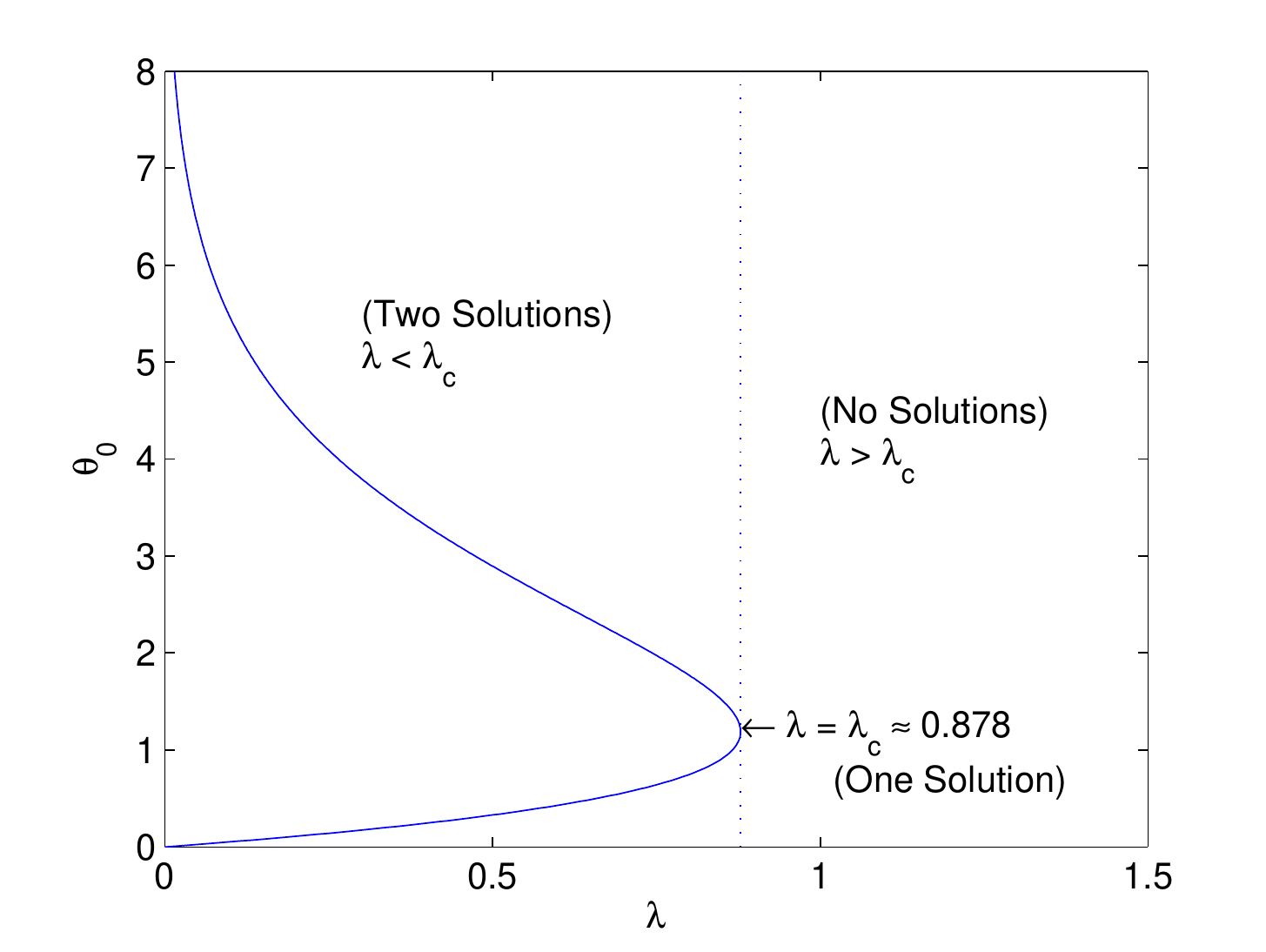}
  \caption{\small Response diagram for $\theta_0$ as a function of $\lambda$.
    As the value of $\lambda$ decreases, the value of $\theta_0$ has
    multiple possibilities that will lead to a steady state solution
    based on the outcomes of Figure \ref{solmaxtheta0}~\cite{fowler}.} 
  \label{lambdavstheta0}
\end{figure}

Define
\begin{equation}
  w=\left(\frac{\lambda}{2}\right)^{1/2}e^{\theta_0/2},
\end{equation}
as well as the functions
\begin{subequations}
  \begin{align}
    y_1&=\cosh w, \\
    y_2&=e^{\theta_0/2}=\left(\frac{2}{\lambda}\right)^{1/2} w.
  \end{align}
\end{subequations}
There exists only one solution if $y_2$ is tangent to $y_1$ at some
point $w=w^*$.  The tangent is given by
\begin{subequations}
  \begin{align}
    \frac{d}{dw}\left(\frac{2}{\lambda}\right)^{1/2} w &= \frac{d}{dw} \cosh w, \\
    \left(\frac{2}{\lambda}\right)^{1/2} &= \sinh w, \label{tangent}
  \end{align}
\end{subequations}
and point $w^*$ is found from
\begin{align}
  w^*&=\left(\frac{\lambda}{2}\right)^{1/2}\cosh w^*
  %=\left(\frac{\lambda}{2}\right)^{1/2}\left(1+\sinh^2 w^*\right)^{1/2}
  =\left(\frac{\lambda}{2}\right)^{1/2}\left(1+\frac{2}{\lambda}\right)^{1/2}
  %=\left(\frac{\lambda}{2}+1\right)^{1/2}
  =\left(\frac{\lambda+2}{2}\right)^{1/2}. \label{w_star}
\end{align}
Upon substitution of (\ref{w_star}) into (\ref{tangent}), the critical
value, $\lambda_c \approx 0.878$, is given by the unique value that
satisfies 
\begin{equation}
  1=\sqrt{\frac{\lambda_c}{2}}\sinh\left(\sqrt{\frac{\lambda_c+2}{2}}\right),
\end{equation}
which was found using Maple.

Figure \ref{lambdavstheta0} gives the value of $\theta_0$ as a
function of $\lambda$.  There exists an asymptote at $\lambda_c$,
where there are 2 solutions for $\lambda<\lambda_c$, 1 solution for
$\lambda=\lambda_c$ or 0 solutions for $\lambda>\lambda_c$.

\subsection{Fickian Diffusion}

In order to find a solution to expression (\ref{phi_l_nondim}), let
$\theta=-\phi_l$ in Eq.~(\ref{fowler_ode}).  Then this system can be
solved analytically using the same approach as the one above.

\begin{figure}[t]
  \centering
  \includegraphics[width=0.8\linewidth]{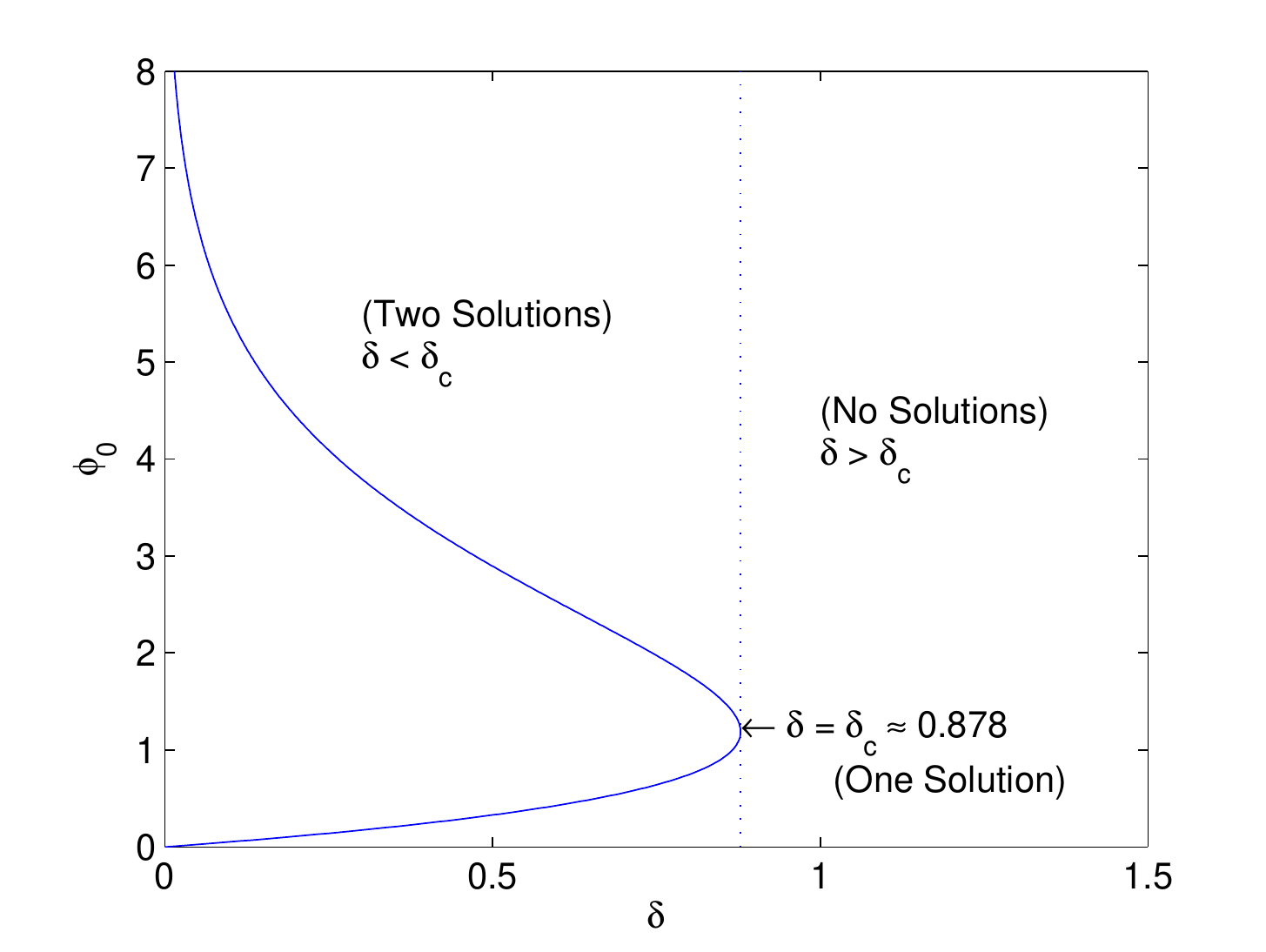}
  \caption{\small Response diagram for $\phi_0$ as a function of $\delta$.  The
    result is the same as for thermal runaway where steady state solutions
    appear when $\delta\leq\delta_c$.} 
  \label{deltavsphi0}
\end{figure}

This time, applying the boundary condition at the right boundary ($\phi(1)=0$)
gives the minimum value $\phi_0$ which satisfies
\begin{equation}
  e^{-\phi_0/2}=\cosh\left(\sqrt{\frac{\delta}{2}}e^{-\phi_0/2}\right).
  \label{minphi}
\end{equation}
Once again, we can find the critical value, $\delta_c \approx 0.878$,
by examining the solutions to Eq.~(\ref{minphi}) in the same way as
Eq.~(\ref{maxtemp}).  Here, $\delta_c$ is given by
\begin{equation}
  1=\sqrt{\frac{\delta_c}{2}}\sinh\left(\sqrt{\frac{\delta_c+2}{2}}\right).
\end{equation}
The solutions are the same and given in Figure \ref{solmaxtheta0}
where $\theta=-\phi$ and $\lambda=\delta$. 

Figure \ref{deltavsphi0} shows the value of $\phi_0$ as a
function of $\delta$.  There exists an asymptote at $\delta_c$,
where there are two solutions for $\delta<\delta_c$, one solution for
$\delta=\delta_c$ or no solutions for $\delta>\delta_c$.  In the case
of two solutions, the ``high-voltage'' solution (i.e., larger value of
$\phi_0$ for given $\delta$) is very likely unstable (without proof).
Therefore, this solution cannot be observed experimentally.

In the MCFC cathode model, as the liquid conductivity decreases,
$\delta$ will increase (see Eq.~(\ref{delta_eq})).  The analytical
results presented here explain the non-existence of steady state
solutions found in the numerical computations.  For large values of
$D$ and $\sigma_s$, resulting in near constant functions $c_{\ms
  O_2},\,c_{\ms CO_2}$ and $\phi_s$, we find very good agreement
between the theoretical value of $\delta_c$ and the value obtained
numerically. 

Also, as $\delta\to\delta_c$, the CPU time increases dramatically.

\section{Optimization of a MCFC Cathode}
\subsection{Full Model Optimization}
Initially, the optimization of the porosity in the cathode was
attempted for the full model of four variables by using the Fickian
Diffusion model with the built-in optimization routines in MATLAB
(i.e., ${\sf fmincon}$).  The gas porosity was taken to be a function of the
position across the cathode, which affects the results through the
Bruggeman term ($\epsilon_i(x)^b$), while the liquid porosity was
updated at each position based upon a constant solid porosity.  The
porosity was updated after convergence to a solution based upon
maximizing the current density at the channel.  The flux at the
channel for the solid potential was used for the maximization
criteria.

Results show that either the model is ill-posed or there exists a
point where the optimization routine can no longer alter the porosity
to reach a maximum for the current.  Each time the program was run for
the same parameters but different, random initial porosities, the
results were inconsistent with each other and were not representative
of what was expected based on the physics of the problem.  At the
channel, it is expected to have a larger porosity than at the
electrode/electrolyte interface for the gas porosity to allow for more
of the gas species to flow into the electrode and react accordingly.
Hence, this optimization approach using ${\sf fmincon}$ was not pursued any
further and a different optimization problem was investigated instead,
employing optimal control methods.
\subsection{Optimal Control}
\label{optimal_control}
Optimal control problems can be used to measure how effective a given control
of a system is by minimizing a cost functional.  This type of problem
is an important part of optimization and has many applications.  The
following is a formulation of optimal control problems from
Pedregal~\cite{pedregal}.  

The state of a given system is described by a number of parameters,
\begin{equation}
x=(x_1,x_2,\ldots,x_n),
\end{equation}
which evolve according to a state equation,
\begin{equation}
x'(t)=f(t,x(t),u(t)),
\label{int_opt_state}
\end{equation}
with boundary or initial conditions 
\begin{align}
x(0)=x_0,&&x(T)=x_T.
\end{align}
The control parameters are given by
\begin{equation}
u=(u_1,u_2,\ldots,u_n),
\end{equation}
and they may depend on $t$ also.

The functional we are attempting to minimize is defined as
\begin{equation}
I(x,u)=\int^T_0 F(t,x(t),u(t))\,dt.
\end{equation}
Both the state equation and the objective functional depend upon the
control parameters $u$.

A pair $(x,u)$ is said to be feasible and admissible if the following
is fulfilled~\cite{pedregal}:
\begin{enumerate}
\item constraints on the control: $u(t)\in K$ for all $t\in (0,T)$,
  where $K$ is the permitted range;
\item state law: $x'(t)=f(t,x(t),u(t))$ for all $t\in (0,T)$;
\item end-point conditions: $x(0)=x_0$, $x(T)=x_T$.
\end{enumerate}
An optimal control problem may consist of having both or only single
endpoint conditions, and transversality conditions may be needed to complete
the formulation of the problem, which will be discussed later.

An admissible pair $(X,U)$ is sought such that,
\begin{equation}
I(X,U)\leq I(x,u)
\end{equation}
for all other feasible pairs $(x,u)$.

We can incorporate the point-wise constraint Eq.~(\ref{int_opt_state}) using
a Lagrange multiplier or co-state $p(t)$ and consider the augmented functional
\begin{equation}
\bar{I}(x,u,p,x')=\int^T_0 \left[
  F(t,x(t),u(t))+p(t)\cdot(f(t,x(t),u(t))-x'(t)) \right]
dt.
\end{equation}
The optimal solutions for the optimal control problem can be found
from the Euler-Lagrange equations for $\bar{I}$.  
\begin{theorem}
  \emph{(Euler-Lagrange Equation)}
  \label{E-L}
  If $x$ is an optimal solution of $\int_{\Omega} F(t,x(t),u(t))\,dt$,
  then $x$ must also be a solution of the problem (E-L)
  \begin{align}
    div(F_{x}(t,x(t),\nabla x(t)))&=F_{\nabla x}(t,x(t),\nabla
    x(t))\,\in\,\Omega, &x&=x_0\,\in\,\partial\Omega,
  \end{align}
  for the variables ($x$,$\nabla x$) with either prescribed or
  transversality conditions applied on $\partial\Omega$.
\end{theorem}
The Euler-Lagrange equations for the optimal control problem can be
found first by defining
\begin{equation}
G(t,u,p,x,u',p',x')=F(t,x,u)+p\cdot(f(t,x,u)-x'),
\label{opt_G_EL}
\end{equation}
where we note that the right-hand side of (\ref{opt_G_EL}) has no
explicit dependence on either $u'$ or $p'$.

Then the system can be written as
\begin{equation}
\frac{d}{dt}\frac{\partial G}{\partial x'}=\frac{\partial G}{\partial
  x}, \hspace{3mm} \frac{d}{dt}\frac{\partial G}{\partial
  u'}=\frac{\partial G}{\partial u}, \hspace{3mm}
  \frac{d}{dt}\frac{\partial G}{\partial p'}=\frac{\partial G}{\partial p},
\end{equation}
and, therefore,
\begin{subequations}
  \begin{align}
    -p'&=\frac{\partial F}{\partial x}(t,x,u)+p\frac{\partial f}{\partial
      x}(t,x,u), \label{el1} \\
    0&=\frac{\partial F}{\partial u}(x,u,t)+p\frac{\partial f}{\partial
      u}(t,x,u), \label{el2}  \\
    0&=x'-f(x,u,t). \label{el3}
  \end{align}
\end{subequations}
Defining the Hamiltonian of the system as $H=F+pf$, we can write
\begin{subequations}
  \begin{align}
    p'&=-\frac{\partial H}{\partial x}, \label{H1}\\
    H(u)&=\min_{v\in K} H(v),\label{H2} \\
    x'&=f(t,x,u). \label{H3}
  \end{align}
\end{subequations}
where $K$ is the set of all admissible controls.

This provides a system of first-order differential equations,
(\ref{H1}) and (\ref{H3}), for which we need boundary conditions,
while (\ref{H2}) is an algebraic constraint, which includes the end
points in the case of a finite domain.  We have at least one
boundary condition for the equation involving $x'$, while the boundary
condition for $p'$ is completed with the transversality condition or
natural boundary condition.
\begin{theorem}
\emph{(Transversality Condition)}
\label{tranversality}
If at a given endpoint (initial or final) we have a condition on the
state, we do not enforce the corresponding transversality condition,
but if the state is free, then the transversality condition $p=0$ at
the given endpoint must be taken into account~\cite{pedregal}.
\end{theorem}
This section provides a framework for an optimal control problem
involving the diffusion of chemical species across an electrode domain.

\subsection{Single Species Optimization}
\label{opt_problem}
In order to build an optimization routine without the use of built-in
functions, a simpler model was introduced that contained only one
differential equation that represents a single species diffusing
across the domain.  Instead of minimizing the current, the objective
function consisted of maximizing the reaction rate balanced with a
function representing the costs and durability of the electrode.  This
type of optimization would ensure longevity for the cathode material
as well as maximizing the current produced.

Fick's Laws were again used to produce the reaction-diffusion
model\footnote{Strictly speaking, we need to solve
  $-\frac{d}{dx}\left(D(x)\frac{dc}{dx}\right)= -\frac{dD}{dx}\frac{dc}{dx}-
  D\frac{d^2c}{dx^2} = -ac$.  Hence, we shall see that for large
  $\alpha$ in (\ref{g_of_D}), the term $-\frac{dD}{dx}\frac{dc}{dx}$
  can be neglected.  As $\alpha$ decreases, the term gains in
  importance.  Future work will include the optimization of the full model
  $-\frac{d}{dx}\left(D(x)\frac{dc}{dx}\right)=-ac$, possibly by
  utilizing the routine ${\sf fmincon}$ in MATLAB or other commercial software.}
\begin{equation}
-D\frac{d^2c}{dx^2}=h(c)=-ac,
\label{opt_sec}
\end{equation}
where $D$ is the control parameter exercised on the system.  This
parameter will vary across the domain but remains constant to first
order and we can avoid taking the derivative with respect to the
position in order to complete the optimization.  The function $h(c)$
represents the reaction rate and $a$ contains other variables of the
system.  Since we are using only one equation, we can represent the
reactivity, $a$, as a single constant. 

\begin{figure}[t]
  \centering
  \includegraphics[width=0.8\linewidth]{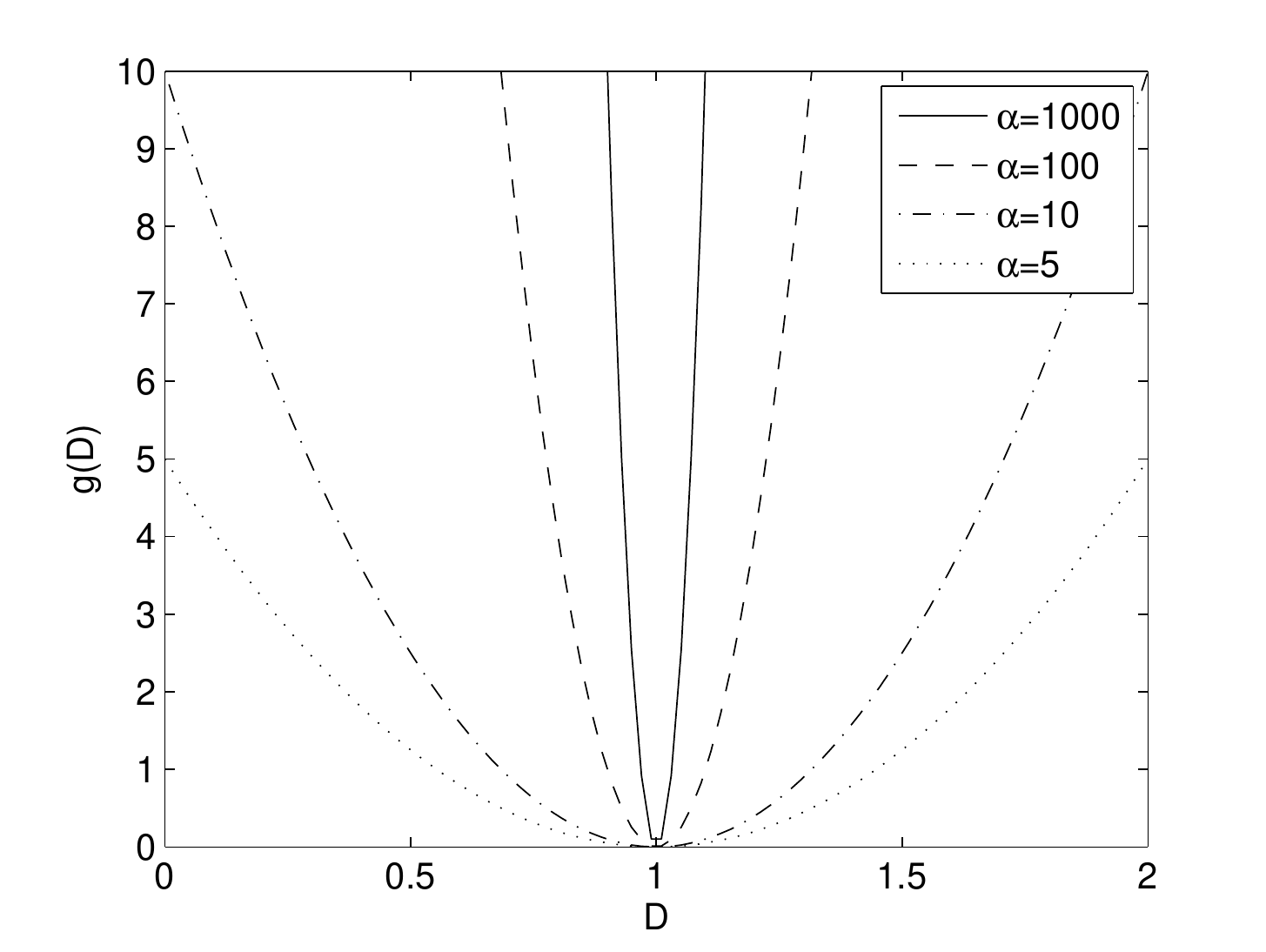}
  \caption{\small Function to be minimized representing the costs and
    durability of the cathode materials.  If the diffusivity, $D$, is
    increased, the porous medium becomes less durable.  If the
    diffusivity is decreased, the pore size becomes much smaller and is
    harder to manufacture for the same volume fraction of species.  In
    order to stay near the optimal value (minimum $D$), the parameter
    $\alpha$ is increased.} 
  \label{g_D}
\end{figure}

The function $g(D)$, shown in Figure \ref{g_D} for different values of
$\alpha$, represents the costs and durability of the cathode material 
\begin{equation}
g(D)=\alpha\left(D-D_0\right)^2.
\label{g_of_D}
\end{equation}
The objective function will measure how good the control is
based upon the criteria discussed above.  It is given by
\begin{equation}
I(c,D)=\int^L_0 \lambda h(c)+(1-\lambda)g(D)\,dx.
\end{equation}
Here, $\lambda$ can be between $0$ and $1$, but chosen to be $0.5$.
If $\lambda$ is $1$, it turns out that the objective function cannot
be minimized as the optimization fails.  If $\lambda$ is $0$, the
objective function is no longer maximizing the reaction rate. 

The parameter $\alpha$ is increased to prevent the diffusion
coefficient from becoming too large or too small in the optimization
process.  If we have a large diffusion coefficient, the porous medium
no longer has an effect on the diffusion across the domain.  It also
means that the electrode is more porous and less stable mechanically.
Note that $D$ cannot exceed the diffusivity in bulk gas.  A small
diffusion coefficient would require smaller pores which are difficult
to manufacture for the same volume fraction of species, resulting in
high costs.  Therefore, as the optimization routine processes, the
diffusion coefficient remains in the vicinity of the minimum value,
$D_0$.  Here, we choose $D_0=1$. 
\begin{figure}[t]
  \centering
  \includegraphics[width=0.8\linewidth]{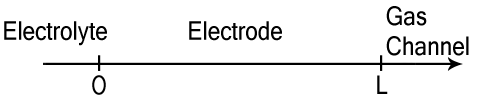}
  \caption{\small Domain schematic for the simplified
  model (\ref{opt_sec})-(\ref{opt_c0}).} 
  \label{opt_domain}
\end{figure}

The boundary conditions are both considered at the origin, which
represents the elec\-trode/elec\-trolyte interface where the gas cannot
flow across
\begin{align}
  c(0)&=c_{\ms 0}, \label{opt_c0}&\frac{dc}{dx}(0)&=0.
\end{align}
As the species diffuses from right to left in the domain (Figure
\ref{opt_domain}), we expect the concentration to reach a specified
value, $c_{\ms 0}$, while maintaining no flux at the left boundary
into the electrolyte.  The channel is considered to be at the right
boundary.  The Dirichlet condition ($c(0)=c_{\ms 0}$), which is typically
found at the channel ($x=L$) is moved to the electrolyte interface so
as to have a well-defined optimization problem.

Using the method described in Section~\ref{optimal_control}, we start
by formulating the second-order differential equation,
Eq.~(\ref{opt_sec}), as two first-order differential equations with
corresponding boundary conditions, 
\begin{align}
  x_1'&=x_2,&x_1(0)&=c_0 \label{opt_fir_u}, \\
  x_2'&=-\frac{h(x_1)}{D},&x_2(0)&=0, \label{opt_fir_v}
\end{align}
where the Hamiltonian is given by,
\begin{equation}
H=\lambda h(x_1)+(1-\lambda)g(D)+px_2-\frac{qh(x_1)}{D}.
\end{equation}
\begin{figure}[t]
  \centering
  \includegraphics[width=0.8\linewidth]{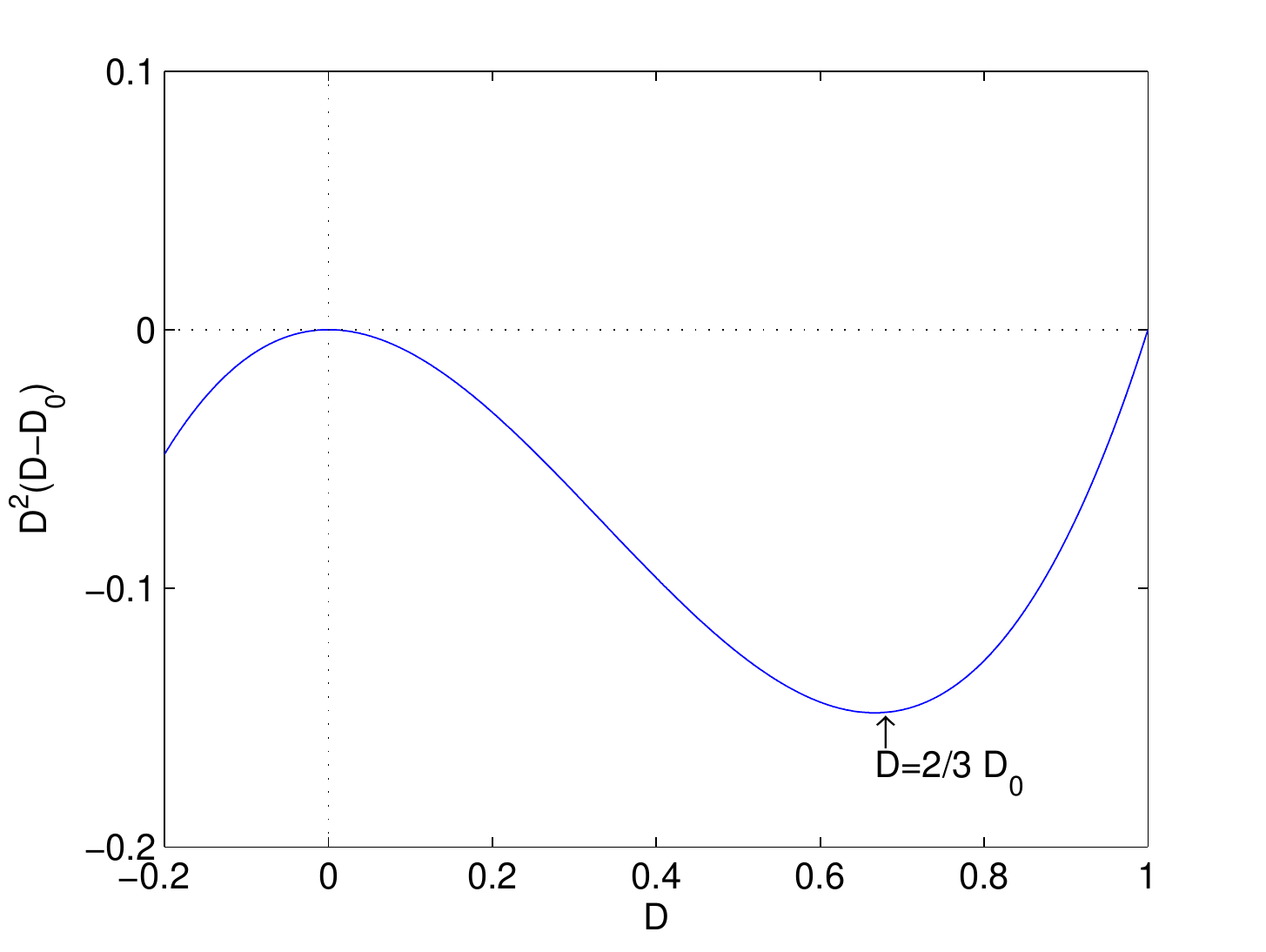}
  \caption{\small The diffusion coefficient will always be a positive-valued
    number, which is determined by Eq.~(\ref{D_quad}).  The smallest
    value of $D$ will occur at the electrode/electrolyte boundary where
    the gas can no longer diffuse.  The minimum value, $\hat{D}$, is
    found where there exists a local minimum in Eq.~(\ref{D_quad}) as
    shown on the graph.} 
  \label{D_hat}
\end{figure}
Here, $p$ and $q$ are co-states of $x_1'$ and $x_2'$.

Using the Hamiltonian, a system of equations is defined based upon the
Euler-Lagrange system, Eqs.~(\ref{H1})-(\ref{H3}),
\begin{subequations}
  \begin{align}
    p'&=-\frac{\partial H}{\partial x_1}=-\left(\lambda h'(x_1) -
    \frac{qh'(x_1)}{D}\right) = -h'(x_1)\left(\lambda-\frac{q}{D}\right), \label{p_prime} \\
    q'&=-\frac{\partial H}{\partial x_2}=-p, \label{q_prime}\\
    H(D)&=\min_{v\in K} H(v)=\min_{v\in K}\,\left[\lambda
    h(x_1)+(1-\lambda)g(v)+px_2-\frac{qh(x_1)}{v}\right], \\
    &\Rightarrow D^2(D-D_0)=\frac{-qh(x_1)}{\alpha},
    \label{D_quad}
  \end{align}
\end{subequations}
where we used $\lambda=1/2$.

The transversality conditions for this problem are given at the
channel,
\begin{equation}
p(L)=q(L)=0.
\end{equation}
The solution will generally require numerical tools.  However, the
problem will first be simplified in order to achieve an analytical
solution of the system of four ODEs, including the algebraic
equation~(\ref{D_quad}).

In our simple model, let $a=c_0=L=D_0=1$ and $\lambda=\frac{1}{2}$.
Then Eqs.~(\ref{opt_fir_u}), (\ref{opt_fir_v}), (\ref{p_prime}) and
(\ref{q_prime}) become
\begin{subequations}
  \begin{align}
    x_1'&=x_2, \label{simp_u} \\
    x_2'&=\frac{x_1}{D}, \label{simp_v} \\
    p'&=\frac{1}{2}-\frac{q}{D}, \label{simp_p} \\
    q'&=-p, \label{simp_q}
  \end{align}
\end{subequations}
with boundary conditions
\begin{align}
  x_1(0)&=1,&x_2(0)&=0,&p(1)&=0,&q(1)&=0.
\end{align}
In Eqs.~(\ref{simp_v}) and~(\ref{simp_p}), $D$ is determined by
Eq.~(\ref{D_quad}), which couples all four equations
(\ref{simp_u})-(\ref{simp_q}).  We see that as $\alpha\to\infty$,
$D\to1$ across the domain since $D=0$ is not permissible.  Setting
$D\equiv 1$ decouples the four equations into 2 pairs of
equations. This is acceptable as a first-order approximation to find
an analytical solution since we will see that this is consistent for
small $\alpha$ as well.  Therefore, we first solve
\begin{subequations}
  \begin{align}
    x_1'&=x_2,&x_1(0)&=1, \\
    x_2'&=x_1,&x_2(0)&=0,
  \end{align}
\end{subequations}
which yields the general solutions
\begin{subequations}
  \begin{align}
    x_1(x)&=C_1\sinh(x)+C_2\cosh(x), \\
    x_2(x)&=C_1\cosh(x)+C_2\sinh(x).
  \end{align}
\end{subequations}
Applying the boundary conditions gives the final solutions
\begin{subequations}
  \begin{align}
%    u(0)=1=C_1\sinh(0)+C_2\cosh(0)=C_2, \\
%    v(0)=0=C_1\cosh(0)+C_2\sinh(0)=C_1, \\
%    \Rightarrow 
    x_1(x)&=\cosh(x), \label{opt_usol} \\
    x_2(x)&=\sinh(x). \label{opt_vsol}
  \end{align}
\end{subequations}
Similarly, we solve the remaining two ODEs
\begin{subequations}
  \begin{align}
    p'&=\frac{1}{2}-q,&p(1)&=0, \\
    q'&=-p,&q(1)&=0.
  \end{align}
\end{subequations}
This gives the following solutions
\begin{subequations}
  \begin{align}
    p(x)&=C_1\sinh(x)+C_2\cosh(x), \\
    q(x)&=\frac{1}{2}-(C_1\cosh(x)+C_2\sinh(x)).
  \end{align}
\end{subequations}
Applying the boundary conditions leads to the two conditions
\begin{subequations}
  \begin{align}
    p(1)&=0=C_1\sinh(1)+C_2\cosh(1),\\
    q(1)&=0=\frac{1}{2}-C_1\cosh(1)-C_2\sinh(1)
  \end{align}
\end{subequations}
with solutions $2C_1 = \cosh(1)$ and $2C_2 = -\sinh(1)$.  Applying
these constants to $p$ and $q$, one finds that
\begin{subequations}
  \begin{align}
    %p(1)=0=C_1\sinh(1)+C_2\cosh(1), \\
    %\Rightarrow C_1=-C_2\frac{\cosh(1)}{\sinh(1)}, \\
    %q(1)=0=\frac{1}{2}-C_1\cosh(1)-C_2\sinh(1), \\
    %\Rightarrow
    %C_2\left(\sinh(1)-\frac{\cosh^2(1)}{\sinh(1)}\right)=\frac{1}{2}, \\
    %\Rightarrow C_2\left(\sinh^2(1)-\cosh^2(1)\right)=\frac{1}{2}\sinh(1), \\
    %\Rightarrow C_2=-\frac{1}{2}\sinh(1), \\
    %\Rightarrow C_1=\frac{1}{2}\cosh(1), \\
    %\Rightarrow 
    p(x)&=\frac{1}{2}\cosh(1)\sinh(x)-\frac{1}{2}\sinh(1)\cosh(x),
    \label{opt_psol} \\
    q(x)&=\frac{1}{2}-\frac{1}{2}\cosh(1)\cosh(x)+\frac{1}{2}\sinh(1)\sinh(x).
    \label{opt_qsol} 
  \end{align}
\end{subequations}
\begin{figure}[t]
  \centering
  \begin{minipage}[t]{0.47\linewidth}
    \includegraphics[width=\linewidth]{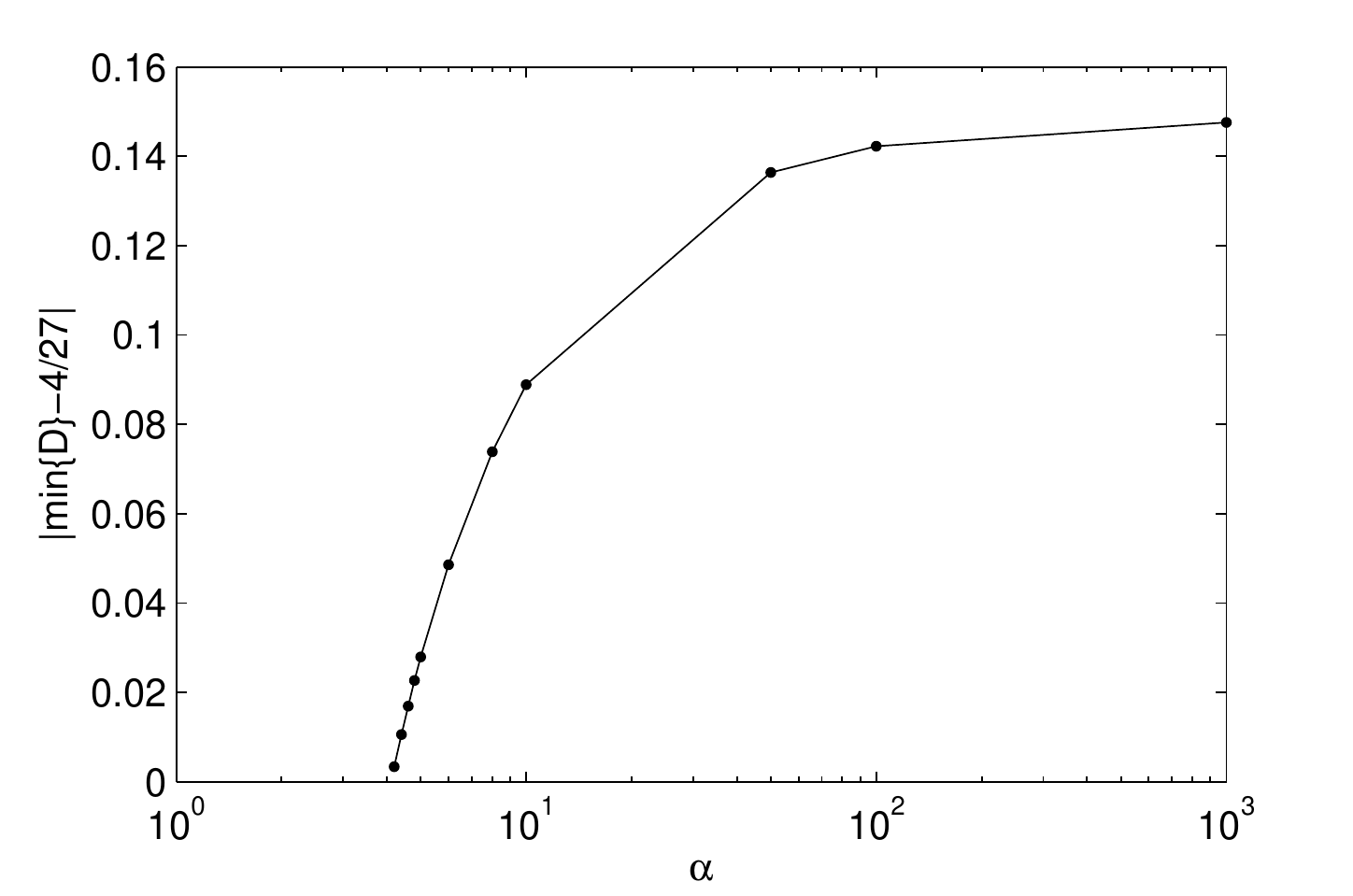}
    \caption{\small As the value of $\alpha$ decreases in $g(D)$, the solution
      begins to diverge as the local minimum of $D$ is passed.} 
    \label{alpha_min}
  \end{minipage}
  \hspace{0.5cm}
  \begin{minipage}[t]{0.47\linewidth}
    \includegraphics[width=\linewidth]{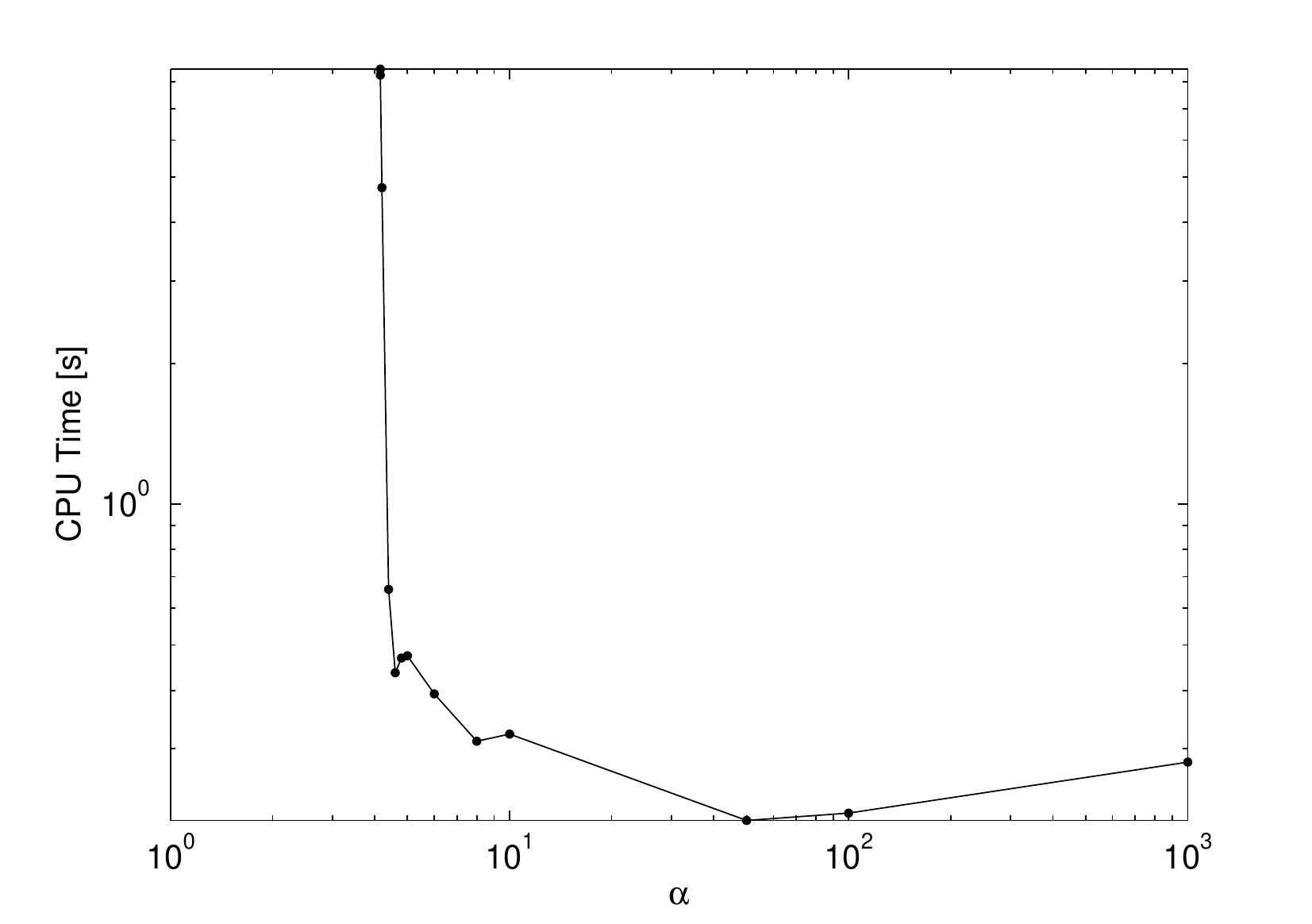}
    \caption{\small As the value of $\alpha_c$ is approached, the CPU time begins
      to increase exponentially as it becomes much more difficult to reach a
      steady state solution numerically.} 
    \label{opt_cputime}
  \end{minipage}
\end{figure}
The solutions to the system of ODEs in the limit $\alpha\to\infty$ are
given in equations~~(\ref{opt_usol})-(\ref{opt_vsol}) and
(\ref{opt_psol})-(\ref{opt_qsol}).

In order to solve this problem numerically for general $\alpha$, a
shooting method is used.  Based upon the analytical results found here
for $\alpha\to\infty$, an initial guess can be obtained from $p(0)$
and $q(0)$ 
\begin{subequations}
  \begin{align}
    p(0)&=-\frac{1}{2}\sinh(1) < 0, &
    q(0)&=\frac{1}{2}\left(1-\cosh(1)\right) < 0.
  \end{align}
\end{subequations}
Hence, we start in the 3rd quadrant of the $(p,q)-\rm{plane}$ and need
to end up at the origin when $x=1$.
\begin{figure}[t]
  \centering
  \includegraphics[width=0.8\linewidth]{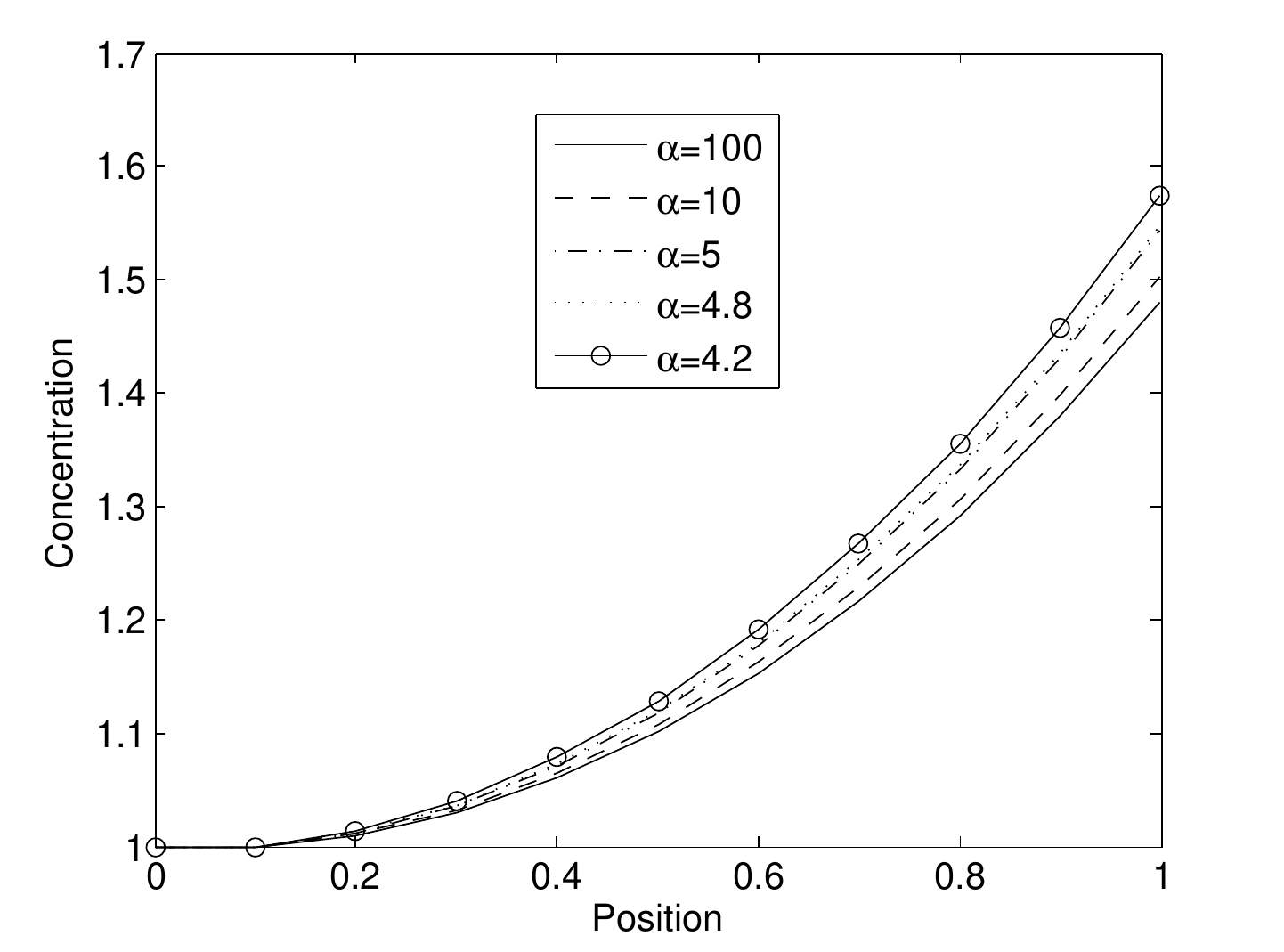}
  \caption{\small The concentration profile across the domain for decreasing
    values of $\alpha$.  As the value of $\alpha_c$ is approached, the
    solution remains within the same order of magnitude.} 
  \label{opt_solution}
\end{figure}

A secant method is used to update the guess for the next iteration for
$\alpha$ which needs two initial guesses.  The second guess is chosen
to be within $1\%$ of the first initial guess so as not to stray away
from the analytical results.  By choosing the solution of the previous
$\alpha$ as the next initial guess, $\alpha$ can be varied and
decreased step-by-step, and the solutions can be found.  Note that we
are strictly shooting in four dimensions, equivalent to a
four-dimensional surface embedded in five dimensions owing to the
algebraic constraint~(\ref{D_quad}).  The algebraic equation is solved
using a built-in MATLAB root solver, ${\sf fzero}$, after every iteration.  

In order for the solution to converge, the right-hand side of
Eq.~(\ref{D_quad}) cannot drop below the local minimum of the
left-hand side or else $D$ would be negative which is unphysical.  The
minimum, $\hat{D}$, is found from, 
\begin{subequations}
  \begin{align}
    \frac{d}{dD}\left(\hat{D}^2(\hat{D}-D_0)\right)&=0,
    \\
    \Rightarrow 3\hat{D}^2-2\hat{D}D_0&=0, \\
    \Rightarrow \hat{D}&=\frac{2}{3}D_0,
  \end{align}
\end{subequations}
where $D_0=1$ is chosen so as to simplify the system.  The value,
$D_0$, will occur at the right boundary since we have $q(1)=0$
in~(\ref{D_quad}).  The diffusivity as a function of $x$ is found
using a root finder at each point, using~(\ref{D_quad}) and the
solutions for $q$ and $x_1$.  The minimum of $\frac{2}{3}D_0$ will
occur close to some critical value of $\alpha$, which is the lowest
possible value where the numerical solution does not exist, and
represents the smallest possible value of $D$ and $\alpha$, as seen in
Figure \ref{D_hat}. The minimum value of $D$ decreases monotonically
with $\alpha$.  The difference between the numerical minimum value,
$\rm{min}\{D\}$, and the analytical value,
$\left(\frac{2}{3}\right)^2\left(\frac{2}{3}-1\right)=-\frac{4}{27}$,
is shown in Figure \ref{alpha_min}.   

Numerically to four decimal places, the critical value of $\alpha$ was
computed as $4.1558$.  As the critical value is approached from
above, the CPU time increases without bound as in
Figure~\ref{opt_cputime}.  As $\alpha\to\alpha_c$, the solution
approaches the graph shown in Figure~\ref{opt_solution}.
\begin{figure}[t]
  \centering
  \includegraphics[width=0.8\linewidth]{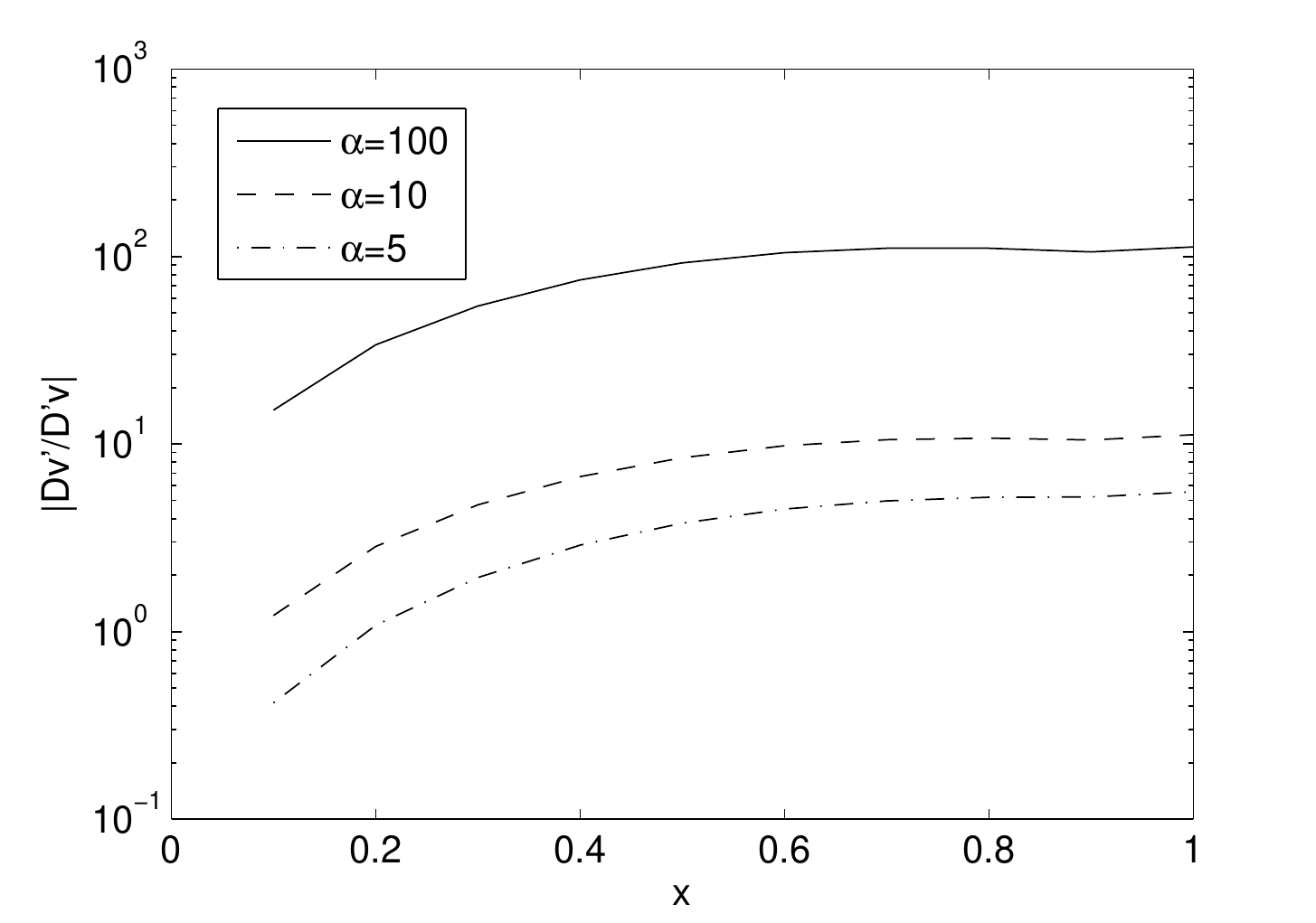}
  \caption{\small Comparison of the two terms on the left-hand side of
   ~(\ref{opt_full_model}).  For large $\alpha$, the ratio is large and
    the model~(\ref{opt_sec}) seems justified.  As $\alpha$ decreases,
    the extra term in the full model~(\ref{opt_full_model}) becomes
    increasingly important.}
  \label{ratio}
\end{figure}

Finally, we would like to gauge whether dropping the term
$-\frac{dD}{dx}\frac{dc}{dx}$ in deriving (\ref{opt_sec}) is
justified.   The presence of this term would modify
Eq.~(\ref{opt_sec}) to give 
\begin{equation}
-\frac{dD}{dx}\frac{dc}{dx}-D\frac{d^2c}{dx^2}=-ac.
\label{opt_full_model}
\end{equation}
For $a=1$, the two terms on the left-hand side can be compared by
their ratio $\left|\frac{Dc''}{D'c'}\right|$.  However, the presence of this
additional term renders our optimization method non-applicable.

Regardless of this fact, we are plotting in Figure \ref{ratio} this
ratio for different values of $\alpha$, approximated by our solution
of the optimization problem $x_1(x),\,x_2(x),\,D(x)$ and so 
\begin{equation}
\left|\frac{Dc''}{D'c'}\right|=\left|\frac{Dx_2'}{D'x_2}\right|.
\label{opt_ratio}
\end{equation}
We see that for large $\alpha$, our original simplification in
deriving (\ref{opt_sec}) seems justified.  However, as
$\alpha\to\alpha_c$ we can observe that the term
$\frac{dD}{dx}\frac{dc}{dx}$ plays an increasingly important role and,
in principle, the full model needs to be optimized,
i.e., Eq.~(\ref{opt_full_model}).

\section{Conclusion and Future Work}
Three different models of diffusion across the cathode of a MCFC based
on Fickian diffusion, convection-diffusion and multi\-component
diffusion with convection were studied.  It has been shown that the
results can differ significantly, depending on system parameters.

The convection-diffusion model shows that for standard values found
from the literature, the convective flux dominates the gas
species' total flux across the domain.  This is a significant
result since this phenomenon has not been taken into account by other
researchers such as White {\em et al}.~\cite{white:2003b} and should be
included in any model of mass transport for a MCFC.  Since the
convective flux is the dominant term, an approximation using only
Fickian diffusion with an effective diffusivity is not possible as
diffusion will only model a decrease in concentration towards the
electrolyte. 

The Maxwell-Stefan equations for multi\-component diffusion take into
consideration the momentum losses to due the interactions between
particles of different species.  Combining it with the convective flux
gives a more detailed view of the mass transport inside the cathode.  For
the standard values from the literature, the results of the
multi\-component model follow those of the simpler convection-diffusion
model with very high accuracy.  In other words, the
convection-diffusion model is a good approximation of the more complex
binary diffusion model.  At low values of the permeability, the
convective flux approaches zero, which allows the differences
between the models to become apparent since the diffusive fluxes do
(simple diffusion) or do not (multi-component diffusion) add up to
zero.

While performing the numerical simulations for values of the liquid
conductivity obtained from the literature, a steady-state solution
could not be achieved for small values.  By using a simpler problem,
an analytical solution was derived showing that there exists a critical
value for the liquid conductivity below which steady-state solutions
do not exist.  This is another significant result as it shows that not
only convection needs to be included in the White {\em et al}.~\cite{white:2003b}
model but the parameter values are inconsistent.

The optimization of the electrode for a fuel cell is of great
importance as it will aid in the manufacturing of cell components to
improve cell performance as well as increase life time.  An analytical
solution has been derived for a simplified system that involves the
optimal profile of the diffusivity across the domain as a control on
the cathode.  The diffusivity can be varied in the more comprehensive
models using the porosity profile across the domain.  Using the data
obtained from the optimization routine, both analytical and numerical,
the porosity may be manufactured so as to improve the performance and
life time of the cell. 

Future work depends upon the availability of parameters that will help
to better understand the convection-diffusion results.  As the fuel
cell begins to reach market acceptance, the optimization of the fuel
cell will grow in importance and will aid in the manufacturing of cell
components to further increase the viability of the MCFC.  A more
realistic optimization of the cathode transport processes needs to be
investigated. 

\cleardoublepage

\section*{Acknowledgement}
We would like to thank Enbridge Inc.\ for their funding and support
through the Ontario Fuel Cell Research and Innovation Network.

\end{document}